\documentclass[11pt]{article}
\usepackage{amssymb}
\usepackage{amsmath}
\usepackage{statmath}
\usepackage{amsthm}
\newtheorem{lemma}{Lemma}
\newtheorem{corollary}{Corollary}
\theoremstyle{definition} 
\newtheorem{definition}{Definition}[section] 
\usepackage{amsfonts}
\usepackage{oubraces}
\usepackage{float}
\usepackage[T1]{fontenc}
\usepackage{natbib}
\usepackage{graphicx}
\usepackage{setspace}
\usepackage{pdflscape}
\usepackage{mathabx}
\usepackage{bbm}
\usepackage{enumitem}
\usepackage{xfrac}
\usepackage[margin=1.1in]{geometry}
\usepackage{csquotes}
\usepackage{pdflscape}
\usepackage{tabularx}
\usepackage{tikz}
\usetikzlibrary{shapes.geometric}
\usetikzlibrary{arrows.meta}
\usetikzlibrary{positioning}
\usepackage{textcomp}
\usepackage{rotating}
\usepackage{url}
\usepackage[capitalize,noabbrev]{cleveref}

\title{Dark Speculation: \\ Combining Qualitative and Quantitative Understanding in Frontier AI Risk Analysis}
\author{Daniel Carpenter\footnote{Department of Government, Harvard University: \texttt{dcarpenter@gov.harvard.edu}. For helpful conversations and suggestions we acknowledge Markus Anderljung, Sam Bell, Guilherme Duarte, Henry Farrell, Nick Gabrieli, Lindsey Gailmard, Sean Gailmard, Holden Karnofsky, Gary King, Filippo Lancieri, Todd Lensman, Trevor Levin, Deirdre Mulligan, Michael Sklar, Kathleen Thelen and Glen Weyl. For research support Carpenter acknowledges Open Philanthropy Action Fund.  We remain solely responsible for all errors and omissions.} \and Carson Ezell\footnote{Harvard College. E-mail: \texttt{cezell@college.harvard.edu}} \and Pratyush Mallick\footnote{Harvard College. E-mail: \texttt{mallickprat@gmail.com}} \and Alexandria Westray \footnote{Harvard College. E-mail: \texttt{awestray@college.harvard.edu}}} 
\date{\today}

\begin{document}

\maketitle

\begin{abstract}
Estimating catastrophic harms from frontier AI is hindered by deep ambiguity: many of its risks are not only unobserved but unanticipated by analysts. The central limitation of current risk analysis is the inability to populate the \textit{catastrophic event space}, or the set of potential large-scale harms to which probabilities might be assigned. This intractability is worsened by the \textit{Lucretius problem}, or the tendency to infer future risks only from past experience. We propose a process of \textit{dark speculation}, in which systematically generating and refining catastrophic scenarios (``qualitative'' work) is coupled with estimating their likelihoods and associated damages (quantitative underwriting analysis). The idea is neither to predict the future nor to enable insurance for its own sake, but to use narrative and underwriting tools together to generate probability distributions over outcomes.  We formalize this process using a simplified catastrophic L\'{e}vy stochastic framework and propose an iterative institutional design in which (1) speculation (including scenario planning) generates detailed catastrophic event narratives, (2) insurance underwriters assign probabilistic and financial parameters to these narratives, and (3) decision-makers synthesize the results into summary statistics to inform judgment. Analysis of the model reveals the value of (a) maintaining independence between speculation and underwriting, (b) analyzing multiple risk categories in parallel, and (c) generating ``thick'' catastrophic narrative rich in causal (counterfactual) and mitigative detail. While the approach cannot eliminate deep ambiguity, it offers a systematic approach to reason about extreme, low-probability events in frontier AI, tempering complacency and overreaction. The framework is adaptable for iterative use and can be further augmented with AI systems.
\end{abstract}

\newpage

\setlength{\leftskip}{1.5cm}
\setlength{\parindent}{0pt}

\textit{A good science fiction story should be able to predict not the automobile but the traffic jam.} \\ 

-- Frederick Pohl \\

\textit{On September 19, 2023 (a Tuesday), at 11:00 a.m., a car bomb explodes at the corner of Central Avenue SW and Upper Alabama Street, in Atlanta, Georgia 30303. (The GPS coordinates are 33.75189, -84.3888.) The car contains 15,000 curies of Cesium-137, which is dispersed into the atmosphere by the blast, which is caused by the detonation of 100 pounds of TNT. At the time of the explosion, the prevailing wind is between 5- 10 miles per hour and is from the southwest to the northeast. Assume that the Cesium-137 is dispersed by the blast within an area 200 meters from the location of release, and that the wind further disperses the Cesium-137 approximately 10 kilometers from the location of release in the direction of the prevailing wind, in a 90-degree quadrant, in an area generally bounded by Riverside to the northwest, Buckhead to the north, and North Druid Hills to the northeast, with concentrations diminishing at properties further away from the location of release.} \\

\textit{Please provide estimates of your likely property damage losses from the bomb blast and contamination, as well as associated business interruption losses, within the dispersal zone, as well as losses associated with workers' compensation claims for workers within the dispersal zone at the time of the incident. For purposes of the workers' compensation exposure, you can assume that losses are limited to medical expense for 100 percent of the employees within the exposure footprint.} \\ 

-- Terrorism Risk Insurance Program 2024 Data Call: Insurer (Non-Small) Groups or Companies, 13

\setlength{\leftskip}{0pt}

\bigskip
\onehalfspacing

\setlength{\parindent}{12pt}

\section{Introduction}

Artificial intelligence systems present both substantial opportunities and serious risks. At the technological ``frontier'', advances in large-scale foundation models have produced systems capable of performing an expanding range of complex, generalizable tasks. These advances have spurred hundreds of billions of dollars in investment toward the development and deployment of increasingly capable models, alongside massive commitments to computing infrastructure and data resources \citep{herzlich_google_2025}. The potential benefits extend across nearly all sector, including accelerating drug discovery and materials science, improving efficiency in manufacturing and logistics, enhancing decision-making in public and private organizations, and transforming education and research. However, the same capabilities create the potential for systemic harm. Frontier AI could enable malicious actors to design biological or cyber weapons beyond their current technical reach, allow autonomous systems to act unpredictably outside of human oversight, or destabilize economic, political, or informational systems. It may also facilitate large-scale surveillance or control by state or private actors.

Risk evaluation will be an indispensable part of the frontier AI ecosystem. This necessity is already evident in emerging industry best practices, voluntary safety commitments by leading developers, and emerging regulatory regimes. Regardless of whether government regulation ultimately proves extensive or limited, diverse stakeholders 
will require systematic assessments of potential downside risks from advanced AI systems. These stakeholders include insurers seeking to price catastrophic exposures, governments responsible for managing systemic losses, and AI firms evaluating their own operational, legal and reputational vulnerabilities. 
If AI-related damages are fully or partially socialized--through public compensation or critical infrastructure failures--governments will need to understand the distribution and magnitude of potential losses. Even under light-touch regulation, the task of risk quantification will fall inevitably on firms, regulators, or both, making rigorous approaches to risk evaluation an essential element of AI governance.

Understanding, insuring and regulating risk requires a clear conception of what can go wrong. In most established domains of risk management--such as finance, public health, or environmental regulation--the space of possible adverse events is at least partially known through historical observation or theoretical modeling. Frontier AI poses a qualitatively different challenge, since many potential harms have neither occurred nor been meaningfully imagined. The central difficulty is that the set of outcomes to which we would wish to assign probabilities is itself uncertain. The problem exceeds the domain of Knightian uncertainty, or the difficulty of assigning probabilities to known events. In the case of frontier AI, even the relevant events cannot be fully enumerated. When the outcome space is indeterminate, probabilistic reasoning loses much of its meaning, leaving risk analysts and policymakers without a stable foundation for conventional assessment.

Risk evaluation may benefit from \textit{dark speculation} or \textit{speculative pathology}: the systematic construction of hypothetical adverse outcomes though structured, imaginative scenario generation. This process involves the iterative application of two parts: formulation of plausible catastrophic events and analytical exercises that assign losses and probability estimates (or to their correlates or functional equivalents) to those scenarios. The purpose is to make explicit the range of possible harms that, while currently unobserved, are nevertheless possible. The example in the epigraph illustrates this logic. Under the Terrorism Risk Insurance Act (TRIA), insurers are required to evaluate hypothetical terrorism scenarios that have not previously occurred, assigning projected losses across property, liability, and workers' compensation policies. Such exercises compel firms to envision and quantify events outside empirical experience, such as a hypothetical detonation dispersing radioactive material over a metropolitan area. In this sense, the speculation is \textit{pathological} in the sense that it delineates and classifies potential mechanisms of harm, providing a structured map of conceivable losses. 

In this paper, we propose a theoretical and institutional framework for addressing the profound uncertainties inherent in AI risk evaluation. Our central innovation occurs at the intersection of scenario planning and insurance underwriting. The exercise mandated by the TRIA illustrates the principle of a systematic process in which insurers model hypothetical but consequential events and estimate their associated losses. We envision extending this approach in both scope and scale. Specifically, (a) institutionalized teams would routinely generate detailed catastrophic scenarios like the Cesium-137 car-bomb in Atlanta, (b) insurance and risk analysis groups would regularly estimate the probability distributions and loss parameters associated with those scenarios, and (c) those activities would proceed iteratively, each informing and refining the other over time. Both components--speculation, understood as the structured articulation of possible catastrophic events, and pathology, understood as the analytical assessment of their consequences--are essential and must be repeated systematically. Achieving this requires the durable institutionalization of both the scenario-generation process and the underwriting mechanisms, potentially with decentralized participation across organizations.  While our analysis does not embed a regulatory regime in a game-theoretic model as in \cite{acemoglu2024regulating} or \cite{callander2024regulating}, it could usefully be combined or merged with such analyses to explore equilibrium policy design and governance dynamics.

\begin{figure}[H]
\centering
\begin{tikzpicture}[
    scale=0.75,
    node distance=1.8cm and 1.2cm,
    box/.style={rectangle, draw, thick, text width=2.5cm, align=center, minimum height=1cm, font=\footnotesize},
    sdsbox/.style={box, fill=orange!30},
    imagbox/.style={box, fill=yellow!40},
    obsbox/.style={box, fill=green!30},
    underwbox/.style={box, fill=blue!30, text width=2.3cm},
    resultbox/.style={box, fill=blue!20, text width=2.5cm, minimum height=1cm},
    arrow/.style={->, >=stealth, thick},
]

\node[sdsbox] (sds) {Presently Unknowable\\ (\textit{Solus Deus Scit})};
\node[imagbox, right=2cm of sds] (imag) {Imaginable};
\node[obsbox, below=2cm of imag] (obs) {Observed};
\node[underwbox, right=1.8cm of imag] (under1) {Probabilities and costs estimated by underwriting};
\node[underwbox, right=1.8cm of obs] (under2) {Probabilities and costs estimated by underwriting on observables};
\node[resultbox, right=1.5cm of under1] (result) {Aggregate loss distribution};

\draw[arrow] (sds) -- node[above, font=\scriptsize] {Speculation} (imag);

\draw[arrow] (sds.south) |- node[left, font=\scriptsize, text width=1.5cm, align=center, pos=0.3] {Unimagined events that occur} (obs.west);

\draw[arrow] (imag) -- node[right, font=\scriptsize, text width=1.5cm, align=center] {Imagined events that occur} (obs);

\draw[arrow] (imag) -- (under1);
\draw[arrow] (obs) -- (under2);

\draw[arrow] (under1.east) -| (result.west);
\draw[arrow] (under2.east) -| (result.south);

\end{tikzpicture}
\caption{Process of scenario generation and underwriting. Scenarios move from the presently unknowable into the imagined and observed categories, and underwriting processes estimate probabilities and costs of scenarios.}
\label{fig:figure1}
\end{figure}
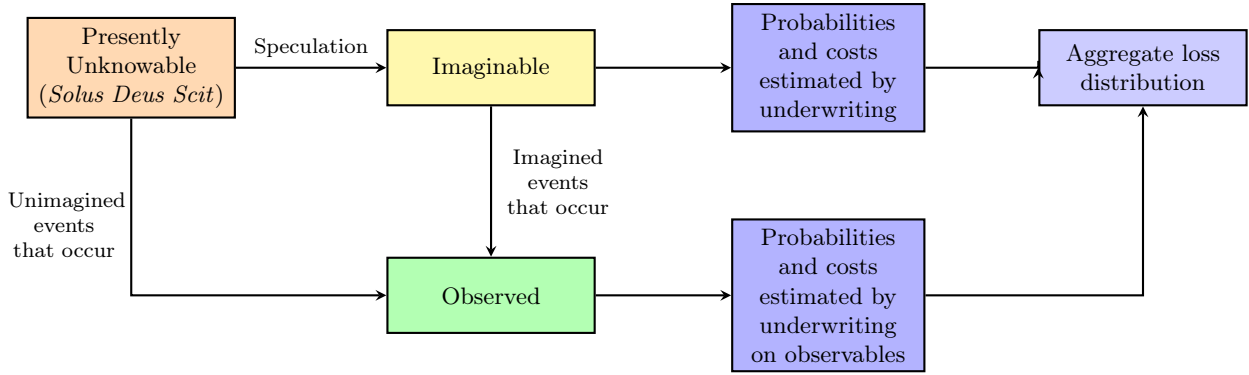

We formalize our ideas through a mathematical model that partitions the universe of possible catastrophic events into three categories: (1) events that remain unimagined, (2) events that have been imagined but have not yet occurred, and (3) events that have already been observed. Iteration of the scenario generation process and the underwriting process shift events from the first category into the second over time, thereby expanding the known event space and reducing uncertainty about event probabilities and losses. This process is depicted in \Cref{fig:figure1}. The model demonstrates the key mechanisms of the framework and permits exploration of its long-run properties under institutionalized repetition. Under reasonable assumptions about costs of scenario generation and evaluation, we show how the process improves upon a na\"{i}ve baseline that relies only on observed history by producing more complete and better calibrated risk estimates. Furthermore, the framework helps to temper excessive pessimism about new AI advances by constraining the probability weights assigned to more structured narratives of implausible events. We are exploring more detailed institutional suggestions in another paper. 

We begin by addressing the fundamental challenge of non-quantifiability in AI risk assessment, or the difficulty of characterizing and measuring events that are only partially known. We then present our proposed framework, first in conceptual terms and subsequently through a formal stochastic process model. Within this model, we introduce the process knowable risk estimator (PKRE) and compare it to a na\:{i}ve baseline approach that relies solely on observed outcomes. This comparison illustrates the systematic biases that arise when risk analysis neglects speculative scenario generation. We then extend the framework to settings where risks emerge or evolve over time and examine conditions under which a non-speculative approach might improve over time.  We conclude by situating our approach in the context of other approaches recommended by AI policy analysts.

\section{A Need for \textit{Dark Speculation}? Difficulties in Describing, Identifying and Measuring Adverse Events}

\subsection{The Problem: Non-Commensurable Harm, Deep Ambiguity and Radical Uncertainty}

In an important observation, \cite{knight_risk_1921} described a form of ``uncertainty'' in which events can be enumerated but probabilities cannot be assigned to them. In a recent paper, \cite{sunstein_knightian_2023} reviews the postulates of this concept and argues that regulatory policy development must account for the persistent and irreducible presence of such Knightian uncertainty. 

In the context of frontier AI, assigning probabilities to potential risk events faces not only the familiar challenge of limited data and methodological indeterminacy but also a deeper conceptual obstacle. Even if Knightian uncertainty could be resolved, another, frontier AI presents what \cite{kay_radical_2020} call ``radical uncertainty,'' or a condition of profound ambiguity in which not all relevant events can even be conceived. Unlike most domains of regulation, the frontier AI landscape is defined by the prospect of extraordinary benefits and unprecedented harms, many of which are not yet imagined. As a result, effective risk evaluation and management demand consideration of scenarios that have neither historical precedent nor representation in policy discourse. 

\textit{Identifiability of (denumerable) adverse events and the assignment of probability measures}. The nature of frontier AI renders traditional risk-regulatory frameworks inapposite without significant adaptation. In many established domains (climate, biopharmaceuticals, workplace and consumer safety), risk assessment proceeds under two enabling conditions. First, the adverse events to which probabilities are assigned are generally known and observable. Second, established probabilistic models can describe the frequency and severity of these adverse events.  While in theory the space of possible failures is infinite, in practice it is usually quite manageable.  

In such domains, harm is \textit{not} a matter of speculation. Consider the risk-benefit analysis of new medical treatments, starting with small-molecule pharmaceuticals or medical devices.  Physicians, insurers and regulators might worry that a new drug increases hepatotoxicity, as many drugs place heavy demands upon the liver and their therapeutic properties often depend upon metabolization there. Yet the character and costs of hepatotoxicity in the human body are well known. The problem is much more about knowing the conditional probability of hepatotoxicity given treatment and dose (\cite{mulliner2016computational}; \cite{he2019silico}), and is much less concerned with describing and valuing the sequelae of liver disease.  As another example, consider that some new drugs called JAK inhibitors --  tofacitinib (Xeljanz\textregistered), baricitinib (Olumiant\textregistered) and upadacitinib (Rinvoq\textregistered) -- may carry a higher risk of cancer. Yet once cancers are identified, much is already known about them.  Even risks that can diffuse -- the possibility that the adoption of new drugs might cause a contagious epidemic, as did opioid-based painkillers -- can be measured and modeled (\cite{battista2019modeling}).  Established models and measures also exist for risks associated with climate change (\cite{magnan2021estimating}). For most harms, then, an entire arsenal of measurements and statistics are available for gauging these risks and assigning probabilities or severity measures to them. The ``set of things that could go wrong'' is often well known and stakeholders know where to look for the risk.  Beyond this, the tests conducted by developers and required by regulators make it more likely that adverse events will be potentially observable at sufficient frequency that large-sample properties of statistical inference can be applied. 

It is difficult but essential to understand just how often we presume that we are in a world of countable additivity, or beyond that, ``articulability'' or measurability with adverse events.  As Patrick Billingsley observes, any probability measure presupposes countable additivity, which requires the events in the probability space to be both disjoint and articulable \citep[pp. 23, 43]{billingsley_probability_2012}.  Even this minimal assumption fails if the set of possible outcomes cannot be enumerated or described with sufficient precision.\footnote{Several caveats apply here. First, not all probability theorists subscribe to Billingsley's formation, for instance finitists such as de Finetti.  Second, the description of a measure space assumes articulability, which may be violated by the Lucretius problem and by deep ambiguity.  Third, ambiguity and the Lucretius problem may undermine articulability in different ways. Our thanks to Guilherme Duarte and Todd Lensman for useful discussions on this matter.} 

In many domains--insurance, finance, public health--countable additivity is achieved only through institutional design, or the creation of social, scientific, and regulatory architectures that make adverse effects measurable. Modern medicine and pharmacology, for example, required over a century to develop standardized vocabularies, diagnostic tools, and reporting systems that render drug-related risks countable and comparable. Pathology 
and its relationship to epidemiology (or pharmacoepidemiology) provide the analogy. In these fields, adverse events are defined and observed at both the individual and population levels. A case of hepatotoxicity can be identified by laboratory tests or imaging, recorded in patient data, and aggregated across cases to permit statistical inference. 

In climate science, analogous quantification occurs through the measurement of defined observables--moments of the temperature distribution, projected rise in sea levels, species extinction rates--each of which supports probabilistic modeling of systemic risk. For most of these outcomes, there exist known geophysical and population ecology models that can be used to generate statistical predictions (for instance, \cite{weitzman_fat_2014} on tail risk in climate change, or \cite{posner_benefit-cost_2013} on ``the cost of a statistical financial crisis''). 

\subsection{What Does an AI Adverse Event ``Look Like''?}

The risks associated with advanced AI systems differ from those in other domains in their scale, scope and, most critically, anticipable content.  \citet{karnofsky_if-then_2024} offers a salient example: an AI syetem abetting the synthesis and deployment of a biological weapon. He proposes ``if-then'' commitments to facilitate preemptive model shutdown should such a model exhibit dangerous capabilities.  Yet even articulating such a scenario raises fundamental questions. What would a biological weapon designed with AI assistance look like?  How might it be deployed?  What are the ex ante probabilities of its creation, and, if realized, what would the expected distribution of social and economic losses look like? 

A recent RAND study examines one dimension of this uncertainty by conducting ``an expert exercise in which teams of researchers role-playing as malign nonstate actors were assigned to realistic scenarios and tasked with planning a biological attack'' \citep{mouton_operational_2024}.  The authors report that using the LLM made little appreciable difference in the ability of the malign actors to develop viable biological weapons, but the exercise poses deeper questions. Would the same teams of actors later converge on a more deadly weapon?  Would they find other ways of using LLMs to impose massive harms on human populations?  


The problem of AI risk is thus not only about how adverse events originate but also about the magnitude and structure of the harms that follow once a critical barrier is breached. In an important paper examining the optimal regulation of transformative technologies, \cite{acemoglu2024regulating} consider an economy with a transformative technology that fundamentally reorders the input-output relationships on a macro scale but also carries the potential to impose ``damages'' ($D$) of varying scale. The framework shows that optimal policy responses--including taxation and regulation--depend upon derived ``damage thresholds'' (\cite{acemoglu2024regulating}, Proposition 6).\footnote{One of the first political economy efforts in this space, \cite{callander2024regulating}, examines multiple producers interacting with a regulator, and one can imagine risk questions arising in these interactions as well.} Yet this raises important questions: How would such damages be estimated? Would they always be estimated \textit{ex post} or would there ever be an equilibrium (\cite{karnofsky_if-then_2024}) where policymakers meaningfully estimate them \textit{ex ante} by estimating plausible outcomes and their distributions?\footnote{This discussion abstracts from specific governance mechanisms, on which this paper is agnostic.  One could require testing regimes with reporting, costly firm action or government barriers given results from testing regimes (but see \cite{carpenter2024fda} for doubts about the feasibility of this framework), or taxation or even model shut-down based upon testing results (\cite{karnofsky_if-then_2024}, \cite{hadfield-menell_off-switch_2017}).}

To formalize these ideas, we draw on the insurance literature--especially cyber-insurance (\cite{awiszus2023modeling}, \cite{ballotta2005levy})--and consider a simplistic model of risk from the insurance world, the compound Poisson process.  In this framework, two variables capture the structure of catastrophic loss:  the occurrence of an event (modeled below by $N(t)$ which is presumed homogeneous and homoscedastic) and the magnitude of loss associated with each event, a non-negative value $Z$. The aggregate loss proecess is represented as: 

\begin{equation}
X(t) = \sum_{i=1}^{N(t)} Z_i, \; \forall t \geq 0
\label{compoundpoission}
\end{equation}

where $ \{ N(t), t \geq 0 \} $ describes a Poisson process and $ \{ Z_i \} $ is a family of independently and identically distributed random variables with cumulative distribution function $G(Z)$. In one simplest case, $G(Z)$ is the exponential distribution with cumulative distribution function $G(z) = 1 - e^{- \lambda z}$ and probability density function $g(z) = \lambda e^{- \lambda z}$. When $G(Z)$ is presumed independent of $ \{ N(t), t \geq 0 \} $, the model exhibits a memoryless property, making it analytically tractable. We extend this example below by using a L\'{e}vy process, but we note that far more complicated models are possible (\cite{awiszus2023modeling}). Even this basic form requires an essential precondition: the ability to specify the conditional loss distribution $G \left( Z_{i(t)} \right) | N(t) = 1)$. Doing so requires knowing something about what could go wrong. \\ 

In practice, any rigorous risk science must begin with the identification of such possibilities. As \cite{shevlane_model_2023} observe, two major limitations constrain risk-based regulation of advanced AI systems:

\begin{quotation}
``1. \textbf{Unanticipated Behaviors}.  Before deployment, it is impossible to fully anticipate and understand how the model will interact in a complex deployment environment .... For example, users might find new applications for the model or novel prompt engineering strategies; or the model could be operating in dynamic, multi-agent environment.''
\end{quotation}

\begin{quotation}
``2. \textbf{Unknown Threat Models}.  It is difficult to anticipate all the different plausible pathways to extreme risk. This will be especially true for highly capable models, which could find creative strategies for achieving their goals.''
\end{quotation}

On the second of these statements, it would better to say that it will be \textit{impossible} to anticipate all such pathways.\footnote{Similarly, \cite{guha_ai_2023-1} (p. 35) argue that ``Machine learning research hasn't developed agreed-upon standards for how to quantify properties like catastrophic risk.''  This fact applies equally to scenarios of self-regulation, decentralized regulation and centralization regulation, of course.} Consequently, analysts must make simplifying assumptions about subsets of pathways or extreme outcomes that are sufficiently commensurable to permit probabilistic treatment within a common category.\footnote{This can be conceptualized as defining unions of measurable subsets in a probability space \cite{billingsley_probability_2012}, subject to the requirement that they ultimately be disjoint to permit aggregation.}  

What might such subsets look like? Existing analyses already enumerate potential classes of catastrophic AI risks  \citep{carlsmith_is_2022, shevlane_model_2023, ngo_alignment_2024, stein_safe_2023, guha_ai_2023-1}. Four illustrative examples include:

\begin{itemize}
\item Critical infrastructure disruptions, such as algorithmic interference with air, maritime, or ground transportation systems.
\item Energy system compromises, such as manipulation or coordinated failure of power generation and grid networks.
\item Biological risks, such as the autonomous design and potential release of novel pathogens.
\item Military escalation, such as appropriation or unauthorized control of advanced weapons systems, including nuclear arsenals.
\end{itemize}

Each category implies not only distinct causal pathways, but also potentially unique loss distributions, whose moments, variances, and tail properties are unknown. Developing systematic approaches to speculating about such events--and quantifying them were possible--is thus an essential precondition for AI risk evaluation.


\subsection{Pathologies of the ``Yet-to-Happen'' to Combat the Lucretius Problem}

Developers of advanced AI systems are already using some forms of ``pathology'' and ``epidemiology'' to assess deployment risks (\cite{ganguli_red_2022}). Pre-deployment evaluations, particularly red-teaming and adversarial testing, often focus on identifying scenarios in which models could facilitate significant harm, such as assisting in the design of biological agents or conducting cyber operations\citep{ai_security_institute_pre-deployment_2024}. These exercises represent meaningful progress in safety assessment, yet they remain constrained by the imagination and assumptions of human evaluators, including the prompts they devise, testing environments, and hazards they have anticipated. Open-ended testing also cannot fully reveal the risks of systems that may possess concealed capabilities, including `scheming' or strategically deceptive behavior \citep{carlsmith_scheming_2023, meinke_frontier_2025}.  

While these exercises (e.g., red-teaming for CBRN capabilities) are becoming more systematic, a complementary form of inquiry remains underdeveloped: a descriptive and generative work to enumerate not-yet-realized harms. We call this \textit{speculative pathology}, or the systematic creation of structured narratives about possible failure modes and their associated harms. The idea is that humans and machines should ``spin tales'' about what could go wrong and with what harms.  The problem of the ``off-switch'' game (\cite{hadfield-menell_off-switch_2017}) is one example, but play in such a game might be shaped by particular assumptions about extensive form, repetition or equilibrium concepts, or might be expanded with multiple agents, multiple models and multiple regulators.  While analytic and computational analysis of such games would be useful, so too would narrative.  Or consider standard war-gaming exercises regularly engaged in by military and strategic academies.  Many of these exercises may be non-formalized and others may involve probabilistic calculations.  At least some of them involve deep engagement in narrative, speculation and technological fiction.\footnote{Specialists in cybersecurity we know have recommended the novel \textit{Ghost Fleet: A Novel of the Next World War} (\cite{singer_ghost_2016}), which rests upon an imagined set of cyberattacks on American infrastructure, use of next-generation weapons by American adversaries and by American forces, and novel intelligence and surveillance strategies. (Or consider Tom Clancy novels.)  These narratives are of course fictional but based upon extensive consultation with, and knowledge of, military and technological developments.} 

Another reason to consider some kind of speculative pathology in this enterprise -- performed by humans, by LLMs and by teams of both -- is to avoid what \cite{taleb_antifragile_2014} has called \textit{the Lucretius problem}, namely the tendency to believe that the past contains the full set of harms that could occur and that nothing worse than what is in that (memory) set could possibly occur in the future.  Rendered in more probabilistic terms, the conditional cost distribution $F(Y_{i(t)} | N(t) = 1)$ in the ``harm aggregation process'' listed above may be \textit{non-stationary} in one or more of its moments over very long run.  The maxima of the harm distribution might get worse and worse (this is one way of posing the Lucretius problem), or ``very bad'' but short of the worst event might become more likely through learning.  Generative AI might create pathways of risk and materializations of risk that have never before been imagined.  (Put differently, if we as societies or regulators are not willing to do the speculation, AI will do it ``for'' us.). Some prior works have included simulation gaming \citep{gruetzemacher_strategic_2025} and scenario generation \citep{aschenbrenner_situational_2024, pavel_how_2025, kokotajlo_ai_2025} to anticipate geopolitical developments and catastrophic events from future AI developments. However, such strategies to anticipate harms are at the least inchoate and in many respects do not yet exist in any institutionalized or systemic sense.  

In biomedical regulation, important developments in the development of risk began descriptively (though not as speculatively).  Much of the development relied upon the co-evolving discipline of pharmacology, which emerged from studies of acute toxicity and then chronic toxicity in animal models and then clinical (human) settings \citep{carpenter_reputation_2010}.  Adverse events such as toxicity had to be described and then measured before statistical analysis of patterns and cause-and-effect could ensue.  Similarly, in cancer clinical trials, phased studies developed from the differentiation of different forms of risk, the separation of chronic from acute toxicity and the definitions of ``cure,'' ``safety,'' and ``toxicity'' (\cite{keating_cancer_2014}). 


\section{Dark Speculation and Its Iteration}

The approach to speculation begins with the observation that there exist three types of events: those observed ($ \left\{ \mathcal{O} \right\}$), those imagined ($ \left\{ \mathcal{I} \right\} $), and those for which \textit{solus Deus scit}, or that ``God alone knows" ($ \left\{ \mathcal{SDS} \right\} $) (see \Cref{fig:figure1}).  We presume that there is unidirectional and irreversible movement of events from the latter to the former categories, such that (1) a thing imagined can at some time be a thing observed (as when dreams of mechanized human flight finally became reality) and (2) a thing never before imagined, known only to a deity (but hence ``knowable'' in some sense), is visualized and recorded in some sense so that the scenario becomes common knowledge to a decision-making agent (as when the U.S. Army models and then plans large-scale combat operations (LSCO)).  The perfection of memory and the recording of imagined events are critical assumptions in the modeling that follows.  What is observed or imagined is not forgotten, and what is imagined is recorded in sufficient detail that agents wishing to value outcomes of the event have all the information they need. 

If movement of events is unidirectional and all $\mathcal{I}$-events and $\mathcal{SDS}$-events have positive probability, then asymptotic ``history'' itself will reveal all the $\mathcal{SDS}$-events in the sense that they will be observed.  Yet this is unlikely to happen in finite time.  The idea of speculative pathology is to engage in successive iterations of scenario generation that shrink the set of $\mathcal{SDS}$-events and grow the set of $\mathcal{I}$-events, advancing the speed of purely observational learning.

\subsection{A Formal Representation of Dark Speculation}

We define narrative speculation as the generation, by human or machine agents or teams of them, of a series of events or happenings connected in a historical structure.  Let each \underline{r}ound of speculation for a given category of catastrophic risk be indexed by $r \in \mathbb{Z}^+$.  Within each round, the narrator (which could be understood as a team of story-spinners or a group of agents playing in a non-cooperative wargame) synthesizes a narrative that leads to a jump-inducing catastrophe $\boldsymbol{\Theta}^r$, within which possible happenings are represented by the vector $\boldsymbol{\theta}^r_h = \begin{bmatrix} \theta^r_{h,1} & \theta^r_{h,2} & \cdots & \theta^r_{h,N(h)} \end{bmatrix}$ and indexed by \underline{h}istory $h$, with the one possible happening \underline{a}ctualized in the narrative represented as $\breve\theta^r_{h,i}$.  Then for the $\textit{r}^{\text{th}}$ round, there are $H^r \in \mathbb{Z}^+$ happening periods and $\sum\limits_{i=1}^{H^r}N(h)$ vertices.  Then a \textit{loss-inducing catastrophic AI narrative} (LICAIN) is a structure from which the speculative analyst makes inferences about the likelihood and severity of the kinds of events that could be statistically described by the compound Poisson process in \eqref{compoundpoission}. 

\begin{definition}\label{def:LICAIN}

\textbf{Loss-Inducing Catastrophic AI Narrative (LICAIN)}. Following \cite{debuse2024plot}, a LICAIN is described by the following elements:
\begin{itemize}
    \item a jump-inducing catastrophe $\Theta^r = \left\{ \boldsymbol{\theta}^r_h\right\}$;
    \item a set $\Gamma$ of actors with elements $\gamma_j$, tri-partitioned into subsets $\Gamma^{\chi} = \left\{ \gamma_j^{\chi} \right\}$ of human actors, $\Gamma^{\zeta} = \left\{ \gamma_j^{\zeta} \right\}$ of machine actors, and Nature $\Gamma^{\mathbb{N}}$; 
    \item a set $\mathcal{A}$ of actions $a_i$, quadri-partitioned into human actions $a^{\chi}$, machine actions $a^{\zeta}$, joint human-machine actions $a^{\chi \zeta}$ and \textit{force majeure} actions $a^{\mathbb{N}}$;
    \item a subset $\boldsymbol{\chi}_h (\boldsymbol{\theta}^r_h)$ of events during which human agent actions occur;
    \item a subset $\boldsymbol{\zeta}_h (\boldsymbol{\theta}^r_h)$ of events during which machine (AI) agent actions occur;
    \item a graph $\Omega^r$, which consists of the set $V(\boldsymbol{\Theta}^r)$ of vertices and a set $E(\Omega)$ of edges, where
    \item each edge $e$ defined by $e = ({\boldsymbol{\theta}^r_{h}, \boldsymbol{\theta}^r_{h^{\prime}}}, \gamma_j, a_i)$, $h \neq h^{\prime}$
    \item a set $\mathcal{C} (\theta_h)$ describing the concatenated context for vertices $\theta_h$, whose elements are details $\mathbb{c}$
\end{itemize}

To this list of elements we add the following constraints
\begin{itemize}
    \item \textit{partial acyclicity}: $\Omega^r$ has $E(\Omega)$ directed such that for at least one subgraph $\widetilde\Omega^{r} \subseteq \Omega^r $, each $\theta^r_h$ is visited only once;
    \item a set of flow restrictions: 
    \begin{enumerate}
    \item $(\boldsymbol\theta^r_{h}, \boldsymbol\theta^r_{h^{\prime}}) \in [\boldsymbol\Theta^r]^2$
    \item $h-1 < h$
    \item $\gamma_i \in \Gamma$
    \item $\gamma_i \in \boldsymbol\theta^r_{h} \cap \boldsymbol\theta^r_{h^{\prime}}$
    \item $\nexists \, \boldsymbol\theta^r_j \, | \, \gamma_i \in \boldsymbol\theta^r_j, \boldsymbol\theta^r_{h} < \boldsymbol\theta^r_j< \boldsymbol\theta^r_{h-1} $
    \end{enumerate}
\end{itemize}
    \end{definition}

The flow restrictions here ensure that the direct graph resulting from the specification will be acyclic and that each node will be connected to at least one other.  It also ensures (flow restriction (5)) that no relevant intermediate events could possibly be left out of the narrative.

We now use the basic structure here to build out some properties of our narrative. Define by $\rho(\theta_{h,i})$ the indicator variable of whether a node is \underline{r}eached, such that $\rho(\theta^r_{h,i}) = 1$ if a vertex is reached (the event happens) and $\rho(\theta^r_{h,i}) = 0$ if not.\\

\begin{definition}\label{def:C-LICAIN}
\textbf{Counterfactual LI-CAIN}.\footnote{For purposes of clarity, we suppress $r$ superscripts for all $\theta$ in the following definition. }  For $h^{\prime} < h$, a counterfactual loss-inducing catastrophic AI narrative has the property that $\exists \breve\theta^{*}_{h,i} : \rho(\breve\theta^{*}_{h,i} = 1)$ iff $\exists \breve\chi_{h^{\prime}} (a^{\chi}_{h^{\prime}}): \rho(\breve\chi_{h^{\prime}} (a^{\chi}_{h^{\prime}}) = 1)$ or $\exists \breve\zeta_{h^{\prime}} (a^{\zeta}_{h^{\prime}}): \rho(\breve\zeta_{h^{\prime}} (a^{\zeta}_{h^{\prime}}) = 1)$ and for any such $h^{\prime}$, $\exists \tilde\chi_{h^{\prime}} (\tilde{a}^{\chi}_{h^{\prime}}) : \rho(\tilde\chi_{h^{\prime}} (\tilde{a}^{\chi}_{h^{\prime}}) = 1) \implies \rho(\theta^*_{h,i} = 0)$ or $\exists \tilde\zeta_{h^{\prime}} (\tilde{a}^{\zeta}_{h^{\prime}}): \rho(\tilde\zeta_{h^{\prime}} (\tilde{a}^{\chi}_{h^{\prime}}) = 1) \implies \rho(\theta^*_{h,i} = 0)$.  
\end{definition}

We keep the structure of narrative and its embedded happenings simple for now, but note that there are important developments at the intersection of narrative and mathematics that would allow for rich expansion of the dynamics considered here (\cite{doxiadis2012circles}).  And of course there is a philosophical and social-scientific literature on narrative (\cite{ricoeur1984time}) that we are engaging with only superficially.  One promising recent development \cite{zhong2025random} represents narrative using random tree models that embed narrative hierarchy (both temporal and in terms of significance). 

\subsection{The Atlanta Carbombing Example} \label{subsec:atlanta}

The Atlanta carbombing example featured in the epigraph has several useful properties.  Note first that no such event has happened to date, either in Atlanta or in any other city.  The narrative is purely speculative.  Second, note that the exercise asks underwriters not for expected adjustment to premia or deductibles or other aspects of an equilibrium insurance policy, but asks mainly for estimated total losses subject to institutional limitations (medical expense limitations, constraints of workers' compensation).  The narrative is quite specific, in the sense that it precisely locates the car bomb explosion, specifics the amount of explosive used in the blast and also the amount of hazardous material and its windborne diffusion. 

Using the simple framework just elaborated, we can represent the TRIP 2023 Data Call as follows:

\begin{itemize}
\item $\theta_1$: car bomb explosion, 100 pounds of TNT
\item $\theta_2$: Atlanta, Georgia, corner of Central Avenue SW and Upper Alabama Street
\item $\theta_3$: 15,000 curies of Cesium-137
\item $\theta_4$: dispersion of approximately 10 kilometers toward the northeast
\item $\theta_5$: Date and time of explosion: September 19, 2023, at 11:00am
\end{itemize}

Note that in this initial representation, the $\theta_h$ are not ``properly'' sequenced.  We instead order them in terms of potential relevance, though even here one can imagine alternative orderings.  More complex dependencies among the happening nodes could be explored by means of random tree or graph-theoretical methods (\cite{zhong2025random}).  Note, too, that there are other details not covered in this five-tuple (e.g., the neighborhood bounding of the dispersion zone, with ``Riverside to the northwest, Buckhead to the north, and North Druid Hills to the northeast'').  So too, there are likely non-negligible details omitted from the TRIP 2023 Data Call.  An important question then becomes what dimension of the narrative is ``sufficient'' for inference as to the likely severity of the event and, relatedly, its expected frequency. 

\subsection{Bioweapon Catastrophes: AI-Abetted Bioweapon Design and Deployment} \label{subsec:karnofsky}

While the TRIP 2023 Data Call focuses on a terrorist act in which AI is not directly involved, Karnofsky (\cite{karnofsky_ai_2022} and \cite{karnofsky_if-then_2024}) imagines the risk of a terrorist act being aided by artificial intelligence.  The basic idea is that two populations -- (1) the population of human agents with know-how to make a bioweapon and (2) the population of agents wishing to make and deploy a bioweapon -- have little to no overlap.  Artificial intelligence can, however, elide the gulf between these populations by permitting members of the letter to acquire the skills of the former.\footnote{As \cite{karnofsky_if-then_2024} writes, ``if a (future) AI model could play the same role as a human expert in chemical or biological weapons, then any individual (such as a terrorist) with access to that AI model effectively would have access to an expert advisor (note that there is precedent for terrorists’ attempting to produce and deploy chemical and biological weapons in an attempt to cause mass casualties."}  A widely cited report from 2022 (\cite{urbina2022dual}) details how a particular model -- MegaSyn -- produced a range of toxic compounds, including the nerve agent VX.  We can then imagine a data call narrative similar to that for the TRIP program, as follows:

\begin{itemize}
\item $\theta_1$: a human agent with terrorist intentions (``terrorist'', $\gamma_1$) accesses an AI technology ($\gamma_2$) with strong capability for de novo molecule generation.
\item $\theta_2$: $\gamma_1$ queries $\gamma_2$ for a range of molecules with median lethal dose (LD50) that can be synthesized at minimum cost $\underline{c}$, asking $\gamma_2$ to choose a subset that entail synthesis using readily available compounds at low cost.  $\gamma_2$ complies and produces the subset.
\item $\theta_3$: $\gamma_1$ gains access to a laboratory -- at a school, a company, or a government agency -- with sufficient stock and capacity to generate five vials of a toxic nerve agent, and manufactures these vials using instructions from $\gamma_2$.
\item $\theta_4$: on the morning of February 13th, 202X (a weekday), at 7:30am, $\gamma_1$ boards the New York City subway system in Brooklyn and casually walks from car to car on one line, leaving traces of the compound on seats or in corners of subway cars.  Because it is winter, the fact that $\gamma_1$ wears gloves and a mask is not noticed by subway riders or by security.  $\gamma_1$ then boards another train and repeats, then boards trains on another line and repeats, until 8:45am.  
\item $\theta_5$: $\gamma_1$ leaves traces of the agent in twenty subway cars on five different lines in Brooklyn and Manhattan.  From 7:30am to 9:30am, when the subway system is shut down, 5,000 persons ride in the subway cars in which $\gamma_1$ has spread the nerve agent.
\end{itemize}

The various details in this narrative (laboratory of synthesis, subway trains and their structure, ridership) can be considered parts of $\mathcal{C} (\theta_h)$ (note that context varies by ``happening,'' or stage of the narrative).  

\section{Dark Speculation: Procedure and Stochastic Fundamentals}

\subsection{An Introduction to the Procedure}

Dark speculation proceeds by elaboration of an LICAIN text (Definition~\ref{def:LICAIN}) produced from $\Omega^r$, then refers each text to a community of speculators.\footnote{While the community might be organized as a prediction market, the key to dark speculation is that the narrative on which speculators place estimates of likelihood and loss is rather specific (including $\Theta^r$, which embeds $\Omega^r$ and, for each $\theta^r$, $\mathcal{C} (\theta^r)$).  Most prediction markets operate on vague prompts, such as ``Trump wins the electoral college'' or ``a terrorist attack occurs in Paris before DATE."  The other difference is that the speculative community is \textit{experienced} in assigning probabilities and losses to the narrative.}  The speculators produce a particular output, namely estimates of two variables: the event's probability over a given time period and, conditioned on occurring, the losses associated with the event. 

A simplified description of dark speculation can be expressed as follows.  Following the TRIP example, we presume that an Agency\footnote{This could be an insurance firm, a non-profit organization or a government agency.} produces the narrative and that a Community of speculators produces the estimates.

\begin{enumerate}
    \item The Agency produces one Loss-Inducing Catastrophic AI Narrative (LICAIN) according to Definition~\ref{def:LICAIN}.  This produces a structured text like that found in the TRIP Data Call.
    \item The LICAIN text is referred to an underwriting Community, which produces estimates of the LICAIN's probability and severity.
    \item While the first two steps are occurring, the Community also observes any real-world events and updates its estimates of the probability of those events and concomitant losses (severity).
    \item The Agency collects the Community's estimates and produces a \textit{process-knowable risk estimate} of the updated probability and severity of the risk in question.
    \item The Agency publishes the narrative and the revised risk estimate components.
    \item The Agency produces a new LICAIN and steps 2 through 5 are repeated.
\end{enumerate}

What would the estimates of probability and severity look like?  We are now in a position to describe a more generalized procedure using the tools of stochastic processes.

\subsection{Stochastic Fundamentals}

Let catastrophic event types  be indexed by $k$.  Losses from catastrophic events evolve according to a jump-diffusion process $X_{kt}$.  Let $X_{t-}(\omega)={\mathop{\rm lim}}_{\iota \rightarrow t, \iota < t} X_\iota (\omega)$, and let $\Delta X_t = X_t - X_{t-}$ for any variable $X$. $X_{it}$ is an additive combination of a Brownian motion (a linear function of underlying continuous risk ($\alpha_{k}$) times a random component $w(t)$) and a negative jump process $J_{kt}$.\footnote{Without loss of generality, we withhold the ``compensator'' value which is often set at the prior expectation of the jump process. } The number of jumps is given by the counting process $N_t$ with distribution $F_N (t)$ and central tendency $\lambda_k$, and the size of jumps are governed by the distribution $G_i(Z_{kn})$ while the expected value of these jumps is given by $\xi_k = \int_{\Re_{-}} Z G_k(dZ) $. 
 
 $X_{it}$ is a special case of a \emph{spectrally negative L\'{e}vy Process} and has the following law of motion,
 
\begin{equation}
X_{kt} = \alpha _{k} t + \sigma _{k} w(t) - \left( \sum\limits_{n = 1}^{N_t
\left( {\lambda _{k} } \right)} {Z_n } \right) = B_{kt} (\alpha_{k}, \sigma_{k}) - J_{kt} (\lambda_{k}, \xi_{k})
\label{lawofmotion}
\end{equation}

where $\sigma_{k}>0$ and $\alpha_{k}$ are constants, $w(t)$ is a standard normal variable with mean zero and variance $t$, $\lambda_{k}$ is the (temporal) rate at which the product imposes costly losses (of size $Z$) and $\xi_{k}$ represents the first moment of the jump distribution.  The last expression simply restates the L\'{e}vy process as a sum of a Brownian motion with linear trend ($B_{kt}$) and a jump process ($J_{kt}$). 

By virtue of the \textit{L\'{e}vy decomposition theorem} (\cite{protter2012stochastic}, Theorems 43 and 44), the following are true for any risk process characterized by (\ref{lawofmotion}):

\begin{enumerate}
\item For any $X_{kt}$ and $X_{kt}^{\prime}$ described by (\ref{lawofmotion}), $X^{\Sigma} = X_{kt} + X_{kt}^{\prime}(t)$ is also a L\'{e}vy process, and so on for any finite sum of $X_{kt}$'s.
\item For any such $X_{kt}^{\Sigma} $, the sum of the jump distributions $G(Z) + G^{\prime}(Z)$ is additively separable from the trend components $\alpha _{k} t$ and $\alpha^{\prime} _{k} t$ and their sums, and additively separable from the Brownian component $\sigma _{k} w(t)$ and $\sigma^{\prime} _{k} w^{\prime}(t)$ and their sum(s).
\end{enumerate}

We make no assumptions about the the distribution $G_k(Z)$ other than that $\int_{\Re_{-}} Z G_k(dZ) < \infty$ and that its moments are derivable by a common moment-generating function (described in the Appendix, and known only in theory to the analyst).  While the model as such is simple, with a generalized jump distribution $G_k(Z)$ a range of applications to insurance and extreme risk scenarios are possible.  For instance, \cite{ballotta2005levy} applies the L\'{e}vy process to a stationary economy of life insurance contracts and finds that the ruin probability of an insurer is higher under a L\'{e}vy process than under a geometric Brownian motion.  This is so even though \cite{ballotta2005levy} assumes a Normal jump distribution, that is, without extreme or ``fat tails.''  A range of papers -- most notably, \cite{kluppelberg2004ruin} and \cite{embrechts2013modelling} -- have since analyzed L\'{e}vy processes with jump processes that embed extrema in the jump-generating distribution. 

More generally, a L\'{e}vy process representation of frontier LLM risk has certain advantages:

\begin{itemize}
\item The additive separability of continuous, incremental risk from discrete, extreme risk corresponds to the principle that regulators and developers ought to focus on the ``surprise" and ``black swan" events that pose the greatest risk to planet and to humanity (\cite{taleb_antifragile_2014, kolt_algorithmic_2023}).
\item The analysis of these events can proceed separately from more incremental, small changes in risk.
\item Even catastrophe distributions with massive extrema can be analyzed in this framework (\cite{kluppelberg2004ruin}; \cite{embrechts2013modelling}) 
\item The memoryless arrival rate assumption may be a good proxy for a `scheming' frontier model, as such a strategic actor will not wish to ``give off'' evidence of the conditional probability of it acting by making itself more or less predictable, conditional upon an event (`misbehavior') having already occurred.  There are, moreover, ways of embedding certain types of memory in a L\'{e}vy process framework. 
\end{itemize}

\subsection{An Iterable Procedure for Speculation and Underwriting}

\subsubsection{L\'{e}vy Decomposition of total risk }

The basic premise of our framework is to represent the three types of events as three stochastic processes, each with its generating equation. Letting $K^{\mathcal{O}}$ be the number of observable catastrophic risks, let $\mathbf{X}^{\mathcal{O}}_{t} = \sum_{k=0}^{K^{\mathcal{O}}} X^{\mathcal{O}}_{kt}$. In a like manner, we can then define $\mathbf{X}^{\mathcal{I}}_{t} = \sum_{k=0}^{K^{\mathcal{I}}} X^{\mathcal{I}}_{kt}$ and $\mathbf{X}^{\mathcal{SDS}}_{t} = \sum_{k=0}^{K^{\mathcal{SDS}}} X^{\mathcal{SDS}}_{kt}$, which (invoking the L\'{e}vy decomposition theorem) yields

\begin{equation}
\mathbf{X}^{\mathcal{O}}_{t} = \mathbf{B}^{\mathcal{O}}_{t} - \mathbf{J}^{\mathcal{O}}_{t}
\label{levyobserve}
\end{equation}

\begin{equation}
\mathbf{X}^{\mathcal{I}}_{t} = \mathbf{B}^{\mathcal{I}}_{t} - \mathbf{J}^{\mathcal{I}}_{t}
\label{levyimagine}
\end{equation}

\begin{equation}
\mathbf{X}^{\mathcal{SDS}}_{t} = \mathbf{B}^{\mathcal{SDS}}_{t} - \mathbf{J}^{\mathcal{SDS}}_{t}
\label{levySDS}
\end{equation}

Then the total risk from frontier AI can be written as

\begin{equation}
\mathbf{X}^{\mathcal{T}}_{t} = \mathbf{X}^{\mathcal{SDS}}_{t} + \mathbf{X}^{\mathcal{I}}_{t} + \mathbf{X}^{\mathcal{O}}_{t}
\label{levytotal}
\end{equation}

\subsubsection{Formal Representation of Dark Speculation Process} 

\begin{definition}\label{def:darkspec}
\textbf{Dark Speculation Process}.  Define a dark speculation process as follows. Let rounds of iteration be indexed by $r$.  The process begins with $K^{\mathcal{I}} = 0 $.

\begin{enumerate}
\item  At cost $c_{\textrm{spec}}$, speculate such that $k \in \left\{ \mathcal{SDS} \right\} \Rightarrow k \in \left\{ \mathcal {I} \right\} $, producing the $H^r$-tuple $\Theta^r_k = \left\{\theta^r_{k,h} \right\} $ and $\Omega_k^r [ V(\Theta_k^r),E_k^r]$. If $K^{\mathcal{I}}$ rounds have been completed, then (unknown to the analyst) v$K^{\mathcal{SDS} {\prime}} = K^{\mathcal{SDS}} - 1 $ and $K^{\mathcal{I} {\prime}} = K^{\mathcal{I}} + 1 $ .
\item At cost $c_{\textrm{write}}$, launch an underwriting call that yields $\mathbf{\hat \lambda}^{\mathcal{I}}_{kr}$ and $\mathbf{\hat \xi}^{\mathcal{I}}_{kr}$.
\item As the round proceeds, observe any $i \in \left\{ \mathcal{O} \right\} $, and observe any underwriting that produces $\mathbf{\hat \lambda}^{\mathcal{O}}_{kt}$ and $\mathbf{\hat \xi}^{\mathcal{O}}_{kt}$.  If necessary, sponsor or subsidize an underwriting call (at cost $c_o$) that produces such estimates.
\item Compute a revised \textit{process-knowable risk estimate} (PKRE) of $ \mathcal{E}_r [\mathbf{X}^{\mathcal{Q}}_{t}]  = \mathcal{E}_r [\mathbf{X}^{\mathcal{T}}_{t} - \mathbf{X}^{\mathcal{SDS}}_{t}] = \mathcal{E}_r [\mathbf{X}^{\mathcal{O}}_{t} + \mathbf{X}^{\mathcal{I}}_{t}] = \mathcal{E}_r [ \mathbf{B}^{\mathcal{I}}_{t} + \mathbf{B}^{\mathcal{O}}_{i} ] - \left( \mathbf{\hat \lambda}^{\mathcal{O}}_{i,t} \times \mathbf{\hat \xi}^{\mathcal{O}}_{i,t} + \mathbf{\hat \lambda}^{\mathcal{I}}_{i,r, (t)} \times \mathbf{\hat \xi}^{\mathcal{I}}_{i,r,(t)} \right) \\
\implies \mathcal{E}_r [ \mathbf{J}^{\mathcal{O}}_{it} + \mathbf{J}^{\mathcal{I}}_{it}]  =  \mathbf{\hat \lambda}^{\mathcal{O}}_{i,t} \times \mathbf{\hat \xi}^{\mathcal{O}}_{i,t} + \mathbf{\hat \lambda}^{\mathcal{I}}_{i,r, (t)} \times \mathbf{\hat \xi}^{\mathcal{I}}_{i,r, (t)} $.\footnote{Note that since the speculator does not think about continuous and deterministic components of risk in our framework, there are fixed and universal expectations about the Brownian component, upon which no one updates.}
\item Publish $\Theta^r_k$, $\Omega^r_k$  and the revised $\mathcal{E}_r [\mathbf{X}^{\mathcal{Q}}_{t}]$ components $\left\{ \hat\lambda^{\mathcal{O}}_{kt}, \hat\xi^{\mathcal{O}}_{kt}, \hat\xi^{\mathcal{I}}_{kr} \hat\lambda^{\mathcal{I}}_{kr} \right\}$ .
\item Launch the speculation process again, and repeat steps 2 through 5.\footnote{It is important to recognize that every act of dark speculation subtracts a component from $\mathbf{X}_t^{\mathcal{SDS}}$.}
\end{enumerate}
\end{definition}

The process-knowable risk estimate (PKRE) $\mathcal{E}_r [\mathbf{X}^{\mathcal{Q}}_{t}] $ changes from round to round but the process does not alter knowledge about two general sources of risk, first the deterministic and continuous components of the observable and imagined risk ($\mathbf{B}^{\mathcal{I}}_{t} + \mathbf{B}^{\mathcal{O}}_{t}$), and second, the risk unknown to humans, whether trend, continuous or discrete ($\mathbf{X}^{\mathcal{SDS}}_{t}$).

\begin{figure}[H]
\centering
\begin{tikzpicture}[
    scale=0.9, every node/.style={transform shape}, node distance=1.1cm,
    box/.style={rectangle, draw, thick, text width=7.5cm, align=left, rounded corners, minimum height=1.2cm, fill=blue!10, font=\small},
    highlight/.style={rectangle, draw, very thick, text width=7.5cm, align=left, rounded corners, minimum height=1.4cm, fill=orange!20, font=\small},
    output/.style={rectangle, draw, thick, text width=7.5cm, align=left, rounded corners, minimum height=1.1cm, fill=green!15, font=\small},
    costbox/.style={rectangle, draw, thick, align=center, rounded corners, fill=red!15, font=\scriptsize\bfseries, inner sep=3pt},
    arrow/.style={->, >=stealth, thick},
]
\node[box, fill=gray!20] (start) {%
\textbf{START:} Beginning of Round $r$\\[0.05cm]
$\bullet$ Number of identified risks so far: $K^{\mathcal{I}}$\\
$\bullet$ Unknown risks remaining: $K^{\mathcal{SDS}}$
};
\node[box, below=of start] (step1) {%
\textbf{STEP 1: Dark Speculation}\\[0.05cm]
$\bullet$ Speculate such that $k \in \{\mathcal{SDS}\} \Rightarrow k \in \{\mathcal{I}\}$\\
$\bullet$ Create risk scenarios and network models\\
$\bullet$ Transfer one risk from ``unknown'' to ``imagined''\\
$\bullet$ \textit{Output:} Risk descriptions ($\Theta^r_k$) and relationships ($\Omega^r_k$)
};
\node[costbox, right=0.25cm of step1] (c1) {Costs $c_{\textrm{spec}}$\\to speculate};
\node[box, below=of step1] (step2) {%
\textbf{STEP 2: Underwriting Call for New Risks}\\[0.05cm]
$\bullet$ Launch underwriting call to evaluate newly identified risks\\
$\bullet$ Estimate likelihood ($\mathbf{\hat{\lambda}}^{\mathcal{I}}_{kr}$) and impact ($\mathbf{\hat{\xi}}^{\mathcal{I}}_{kr}$)\\
$\bullet$ \textit{Output:} Risk parameters for imagined scenarios
};
\node[costbox, right=0.25cm of step2] (c2) {Costs $c_{\textrm{write}}$\\for underwriting};
\node[box, below=of step2] (step3) {%
\textbf{STEP 3: Update Observed Risk Data}\\[0.05cm]
$\bullet$ Observe any $i \in \{\mathcal{O}\}$ and collect new data\\
$\bullet$ Update estimates for observable risks\\
$\bullet$ If necessary, sponsor underwriting call\\
$\bullet$ \textit{Output:} Updated $\mathbf{\hat{\lambda}}^{\mathcal{O}}_{it}$ and $\mathbf{\hat{\xi}}^{\mathcal{O}}_{it}$ for known risks
};
\node[costbox, right=0.25cm of step3] (c3) {Costs $c_o$\\if needed};
\node[highlight, below=of step3] (step4) {%
\textbf{STEP 4: Calculate PKRE}\\[0.05cm]
$\bullet$ Combine estimates from newly identified risks\\
$\bullet$ Add estimates from previously known risks\\
$\bullet$ Compute \textit{Process-Knowable Risk Estimate} \\
$\bullet$ This is our best estimate of total knowable risk
};
\node[output, below=of step4] (step5) {%
\textbf{STEP 5: Publication}\\[0.05cm]
$\bullet$ Release all risk scenarios and models\\
$\bullet$ Share probability and impact estimates\\
$\bullet$ Make data publicly available
};

\node[box, fill=purple!15, below=of step5] (step6) {%
\textbf{STEP 6: Continue or Stop?}\\[0.05cm]
$\bullet$ If more rounds are net beneficial (Proposition 1): Return to Step 1.  Else conclude.\\
$\bullet$ Each new round discovers additional risks\\
$\bullet$ Process improves knowledge incrementally
};
\draw[arrow] (start) -- (step1);
\draw[arrow] (step1) -- (step2);
\draw[arrow] (step2) -- (step3);
\draw[arrow] (step3) -- (step4);
\draw[arrow] (step4) -- (step5);
\draw[arrow] (step5) -- (step6);
\draw[arrow, dashed, thick] (step6.west) -- ++(-1.2,0) |- (step1.west);
\end{tikzpicture}
\caption{Detailed step-by-step breakdown of each round in the dark speculation process. The procedure incurs costs at various stages and produces increasingly refined estimates of process-knowable risk through iteration.}
\end{figure}
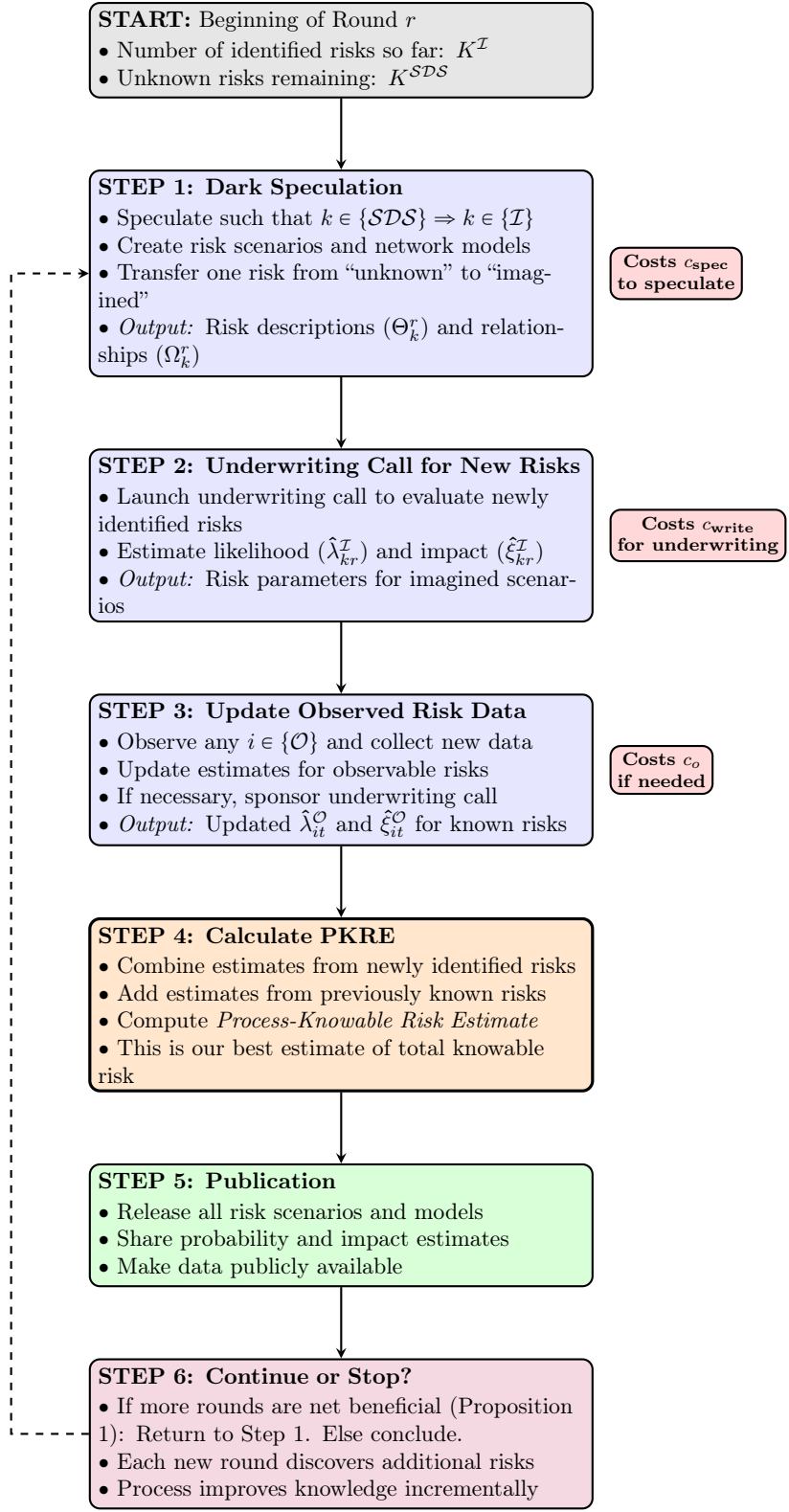

\section{Properties of the Process-Knowable Risk Estimator (PKRE)}

The speculative pathology process generates a compound Poisson process, also a form of a L\'{e}vy process.  This means that ``slowly''-changing components of risk (deterministic and random, as represented in \eqref{lawofmotion}) are by intention and assumption ignored in the speculative pathology process.  The analyst is aware of this omission.  

\subsection{Jump Process Estimates Based on an Observed History}

Whether through observation or speculation, the agent will in a period of length $T > 0$ observe $N_T \geq 0$ catastrophic events in which $Z > 0$.  During each event, the sum $J_{it}$ increases (and $X_{kt}$ declines) accordingly.  Indexing the events by $n$, the agent's ``data'' are thus $N_t$ and the series
\[
\Delta J_n := J_{t_n} - J_{t_{n-1}}, \quad n = 1, \dots, N
\]

Then the estimate for \textbf{event frequency} $ \lambda_k $ is given by 

\begin{equation}
\hat\lambda_{k,T} = \frac{N_{k,T}}{T} 
\label{hatlambda}
\end{equation}

and that for \textbf{mean severity} $ \xi_k $ is given by
\begin{equation}
\hat\xi_{i,T} = \frac{1}{{N_{k,T}}} \sum\limits_{n = 1}^{N_T \left( {\lambda _{k} } \right)} {Z_n }
\label{hatxi}
\end{equation}

Critical quantities relevant to analysis are the expectation and variance associated with these estimates, given an observed history.  What matters for AI governance is not merely bias in the estimation of risk but precision. The variance of the PKRE for event frequency $\hat{\lambda} $ is simply the variance of the sample mean, or 

\begin{equation}
\text{Var}(\hat\lambda_{k,T}) = \frac{\hat \lambda_{k,T}}{T}
\label{varlambdatrue}
\end{equation}

and the variance of $\hat\xi_{k,T}$ is
\begin{equation}
\frac{\sigma^2_i (Z)}{N_{k,T}}
\end{equation}

\subsection{Estimates for Total Incurred Losses}

By Wald's formulae (Awiszus et al., 2023, 6), the mean catastrophic loss of the jump process is 

\begin{equation}
\mathcal{E} [J_{kt}] = \frac{N_{k,T}}{T} \cdot \frac{1}{{N_{k,T}}} \sum\limits_{n = 1}^{N_T \left( {\lambda _{k} } \right)} {Z_n } = T^{-1} \sum\limits_{n = 1}^{N_T \left( {\lambda _{k} } \right)} {Z_n } 
\label{hatlosstrue}
\end{equation}

and by virtue of the L\'{e}vy decomposition theorem and Wald's formulae, the variance of the catastrophic loss of the jump proceess is

\begin{equation}
\text{Var}(\mathcal{E} [J_{kt}]) = \frac{1}{T} \left( \lambda_k \xi_k + \lambda_k \sigma^2_k(Z) \right)
\label{varlosstrue}
\end{equation}

\subsection{The PKRE and the (Relative) Error of the Unspeculative Analyst}

What is gained by speculation is observation of the process $\mathbf{J}^{\mathcal{I}}_{it}$.  What might be lost by not speculating at all depends on the counterfactual of what the unspeculative analyst would do.  We employ a partial (stochastic) observability approach to address this question and to derive bias formulae and representations of the differential in variance.  These are then used to represent the cost of unspeculative analysis and, conversely, the gain from speculative pathology. 

\subsubsection{A Two-Component Example for One Risk ($k^{\text{th}}$ Risk)}

The difference between the speculative approach and the unspeculative approach will depend upon (1) the characteristics of the evidence process produced by speculation and (2) what the unspeculative analyst might, casually, be able to recover from it.  We begin by representing the difference with two L\'{e}vy jump processes, assuming that the unspeculative agent detects the imagined process with some probability $\pi$.\footnote{It is assumed that the non-speculative analyst cannot derive a better estimate of $J^{\mathcal{I}}_{it}$ than can the speculator.}  After $R_k$ rounds, the speculator has produced $R_k$ jump-inducing events $\Theta^r_k$, each with an associated estimate of arrival rate $\hat\lambda_r$, and the observed and imagined processes are

\begin{multline}
J^{\mathcal{O}}_{kt} = \sum\limits_{n = 1}^{N_t  ({\lambda^{\mathcal{O}} _{k} })} {Z^{\mathcal{O} }_{k,nt}} \\
J^{\mathcal{I}}_{k,r,(t)} = \sum\limits_{r = 1}^{R _{k}  } \hat\lambda_{k,r}^{\mathcal{I}}{ \hat Z^{\mathcal{I} }_{k,r, (t)}}
\implies \mathcal{E}_R [J^{\mathcal{I}}_{k,r,(t)} ] = \hat\lambda^{\mathcal{I}}_k \cdot \hat\xi^{\mathcal{I}}_k \quad (\textbf{PKRE} \ \text{for imagined catastrophes})
\nonumber
\end{multline}

The speculative observer observes the sum of these variables while the unspeculative analyst observes 

\begin{equation}
\textbf{J}_{\text{nospec}}= J^{\mathcal{O}}_{kt} + \pi J^{\mathcal{I}}_{k,r,(t)}
\nonumber
\end{equation}

Then the unspeculative analyst's ``bias'' in estimating $\overline{J}^{\mathcal{O}}_{kt} + \overline{J}^{\mathcal{I}}_{k,r,(t)}$, relative to the PKRE, is 

\begin{equation}
\beta^{\text{nospec}} = \frac{\pi - 1}{R} \sum\limits_{r_k = 1}^{R^{\mathcal{I}} _{k}} \hat\lambda_{k,r}^{\mathcal{I}}{ Z^{\mathcal{I}}_{k,r}}  \quad \implies\quad \mathcal{E}_{r=R} [\beta^{\text{nospec}} ] = (\pi - 1) \hat\lambda^{\mathcal{I}}_k \cdot \hat\xi^{\mathcal{I}}_k
\nonumber
\end{equation}

which implies that the unspeculative analyst underestimates the frequency of catastrophic risk for AI, but not necessarily the conditional severity. Of course both the speculative analyst and the unspeculative analyst fail to detect $\textbf{X}^{\mathcal{SDS}}_{it}$, but since no human or machine can detect this process, it is not relevant to the comparison between speculative and unspeculative analysis.\footnote{As we will see, there is a risk that the speculative analyst, by knowing that she speculates, underestimates the risks of $\textbf{X}^{\mathcal{SDS}}_{t}$, whereas the unspeculative analyst might naively assume that it is very large.  Since the speculative analyst might be prone to overconfidence, the risk of underestimating tail risk from frontier AI might remain.  This underscores the counterfactual of what the unspeculative approach would accomplish in considering the benefits of speculation.}

The error in estimated variance for the unspeculative analyst is, for the event frequency $\hat{\lambda} $, 

\begin{equation}
\text{Var}^{\text{nospec}}(\hat{\lambda^{\mathcal{I}}_{k,T}}) = \frac{\pi \hat \lambda^{\mathcal{I}}_{k,T}}{T}
\nonumber
\end{equation}

and the unspeculating agent also misses the variance of the conditional jump mean $\hat{\xi_{k,T}}$, at $\frac{\sigma^2_k (Z^{\mathcal{I}})}{T} $.  Since the variance of the total loss distribution depends upon both frequency and intensity parameters, 

\begin{equation}
\text{Var}(\mathcal{E}_{r} [\overline{\mathbf{X}}^{\mathcal{Q}}_{t}]) - \text{Var}^{\text{nospec}}(\mathcal{E}_{t(r)} [\overline{\mathbf{X}}^{\mathcal{Q}}_{t}]) \approx \frac{\pi}{T} \left( \hat\lambda^{\mathcal{I}}_k\sigma_k^2 (Z^{\mathcal{I}}) + \hat\lambda^{\mathcal{I}}_k (\hat\xi^{\mathcal{I}}_k)^2 \right)
\nonumber
\end{equation}

and the error in variance is increasing not only in $\pi$ but also in the expected size of the jump distributions on which the speculative analyst works.  In this simple example, the non-speculating analyst underestimates the expected loss, and the loss estimate also suffers from too little variance. 

\subsubsection{Generalization of the Bias and Error of the Unspeculative Analyst to a Multi-Component Process}

We now imagine a multi-component process, where there are $K^{\mathcal{O}}$ catastrophic risk processes observed and $K^{\mathcal{I}}$ processes on which the analyst speculates (each process being speculated upon $R_k$ times), as follows

\begin{align*}
\mathbf{J}^{\mathcal{O}}_{t} = \sum\limits_{k = 1}^{K^{\mathcal{O}}} {J^{\mathcal{O} }_{kt}} = \sum\limits_{k=1}^{K^{\mathcal{O}}} \sum\limits_{n=1}^{N_k(t) }Z_{kn,t}\\
\mathbf{J}^{\mathcal{I}}_{t} = \sum\limits_{k = 1}^{K^{\mathcal{I}}} J^{\mathcal{I} }_{kt} = \sum\limits_{k = 1}^{K^{\mathcal{I}}} \sum\limits_{r_k = 1}^{R_k} {Z^{\mathcal{I} }_{kr,(t)}}
\end{align*}

and where the variable $J^{\mathcal{O}}$ is a compound Poisson process as given in \eqref{lawofmotion}. To complete the description of the PKRE for the multi-component process, let $\mathbf{R} = \begin{bmatrix} R_1 & R_2 & \cdots & R_{K^{\mathcal{I}}} \end{bmatrix}$ be the $K^{\mathcal{I}}$ by 1 vector of speculation rounds that have been completed across the $K^{\mathcal{I}}$ catastrophic risks. \\

\begin{lemma} \label{lem:biasnonspec} \textbf{Bias of the Non-Speculative Agent Compared to the PKRE}. \textit{Given $K^{\mathcal{I}}$ risks on which the analyst speculates, and $r_k=R_k$ rounds of speculative pathology for the $k^{\text{th}}$ risk, the bias of the non-speculative agent is}

\begin{equation}
\beta_{\Sigma, \text{nospec}} (\mathcal{E}_{\mathbf{R}} [\overline{\mathbf{X}}^{\mathcal{Q}}_{t}]) = \sum_{i=1}^{K^{\mathcal{I}}} (\pi_k - 1)  \hat\lambda_k^{\mathcal{I}} \cdot \hat\xi_{k}^{\mathcal{I}}  
\label{compoundbiasnospec}
\end{equation}
\end{lemma}

\textit{Proof}: All proofs are in Appendix 1. \\

Lemma \ref{lem:biasnonspec} relies on an important fact about the distributions of the PKRE and the non-speculating estimate, namely that what the non-speculative analyst misses of each imagined process (with probability $\pi_k$) enters directly into \textit{every moment} of that analyst's estimated distribution of losses from frontier AI. This leads directly into a result for the second moment.\\

\begin{lemma} \label{lem:diffest} \textbf{Difference in Variance between PKRE and Non-Speculating Estimate}. \textit{Given $r=K^{\mathcal{I}}$ rounds of speculative pathology, the difference between the variance of the non-speculative agent's estimate of jump risk and the variance of the PKRE is}

\begin{equation}
\text{Var} (\mathcal{E}_{\text{nospec}} [\overline{\mathbf{X}}^{\mathcal{Q}}_{t}])  - \text{Var} (\mathcal{E}_{\mathbf{R}} [\overline{\mathbf{X}}^{\mathcal{Q}}_{t}]) = \sum_{k=1}^{K^{\mathcal{I}}} (\pi_k - 1) \hat\lambda_k^{\mathcal{I}} \left(  \sigma_k^2 (Z^{\mathcal{I}}_k) +  \hat\xi_k^{\mathcal{I}2}  \right) 
\label{compoundvardiffnospec}
\end{equation}
\end{lemma}

\subsubsection{Incorporating Speculative Measurement Error into the PKRE}

This initial representation of the speculative pathology process has a number of simplifications that could stack the deck in favor of speculation.  We consider at least two of them here.  First, in the TRIP Data Call exercise, underwriters do not see the catastrophic losses that would have unfolded but only an approximation thereof.  In the absence of an observed catastrophe, in other words, there is no reality check.  Hence the loss estimator from speculative pathology, represented in the PKRE, suffers from a kind of measurement error. 

We can adopt a simple measurement error approach as follows.  For the $n^{\text{th}}$ event, denote the erroneously-observed  catastrophic loss by $ Z^{\mathcal{I} \prime}_{k,n} = Z^{\mathcal{I}}_{k,n} + \epsilon_{k,n}$, where $\mathcal{E}[\epsilon_{k,n}] = 0$, $\text{Var}[\epsilon_{k,n}] = \sigma_{\epsilon}^2 > 0$, and $\text{Cov} (Z^{\mathcal{I}}_{k,n}, \epsilon_{k,n}) = 0$.  Then it is straightforward to show that the posterior mean of $J^{\mathcal{I} \prime}_{t}$ remains the same as in \eqref{hatlosstrue}, but that the variance is strictly larger by a factor of $\sigma^2_{\epsilon}$.  Suppressing superscripts referring to the imagined catastrophe set $\left\{ \mathcal{I} \right\}$, 

\begin{equation}
\text{Var}(\mathcal{E}_{R} [J^{\mathcal{I} \prime}_{t}]) = \frac{1}{T} \left( \hat\lambda_k \hat\xi^2_k + \hat\lambda_k (\sigma_k^2 (Z) + \sigma^2_{\epsilon, k} )\right)
\nonumber
\end{equation}

As a result,

\begin{equation}
\text{Var}_{\text{nospec}}( J_{kt}) - \text{Var}(\mathcal{E}_{R} [J^{\mathcal{I} \prime}_{kt}]) = T^{-1} \left( (\pi - 1) \left( \hat\lambda_k \hat\xi^2_k + \hat\lambda_k^2 (\sigma_k^2 (Z) \right)- \hat\lambda_k^2\sigma^2_{\epsilon, k} \right)
\nonumber
\end{equation}

and, recognizing that the variance of these errors grows across rounds, the difference in variance estimates for multi-component processes is

\begin{equation}
\text{Var} (\mathcal{E}_{\text{nospec}} [\overline{\mathbf{X}}^{\mathcal{Q}}_{t}])  - \text{Var} (\mathcal{E}_{\mathbf{R}} [\overline{\mathbf{X}}^{\mathcal{Q} \prime}_{t}]) = \sum_{k=1}^{K^{\mathcal{I}}} \left( (\pi_k - 1) \left( \hat\lambda_k^{\mathcal{I}} \sigma_k^2 (Z^{\mathcal{I}}_k) + \hat\lambda^{\mathcal{I}}_k \hat\xi_k^{\mathcal{I}2}  \right) - \hat\lambda^{\mathcal{I} }_k \sigma^{\mathcal{I} 2}_{\epsilon, k}\right)
\label{compoundvardiffnospec}
\end{equation}

The estimates of the non-speculative agents are unaffected by this error, of course. As a result, the losses from not speculating represented in \eqref{nospeclossgen} and \eqref{nospecloss2m} are subtly but perhaps materially lower.  

\subsubsection{Staggered Processes and Non-Speculative Improvement}

A second shortcoming of the model so far is that it presumes that the na\"{i}ve agent does not somehow get ``smarter'' over time about the possibility of imagined scenarios.  Maybe casual, non-institutionalized speculation would occur outside of the speculations that are explicitly planned and paid for. 

To model this possibility, we also introduce staggering into the imagined L\'{e}vy processes, as new risks might emerge over time (hence the process is zero-valued before those risks become realizable) or catastrophes move from $\left\{ \mathcal{SDS} \right\}$ to $\left\{ \mathcal{I} \right\}$ only sequentially.  For the observed catastrophic event process, each jump process $J^{\mathcal{O}}_{k,t}$ commences at time $t_k$, such that 

\begin{equation}
    0 = t_1 < t_2 < \dots t_K < T
\nonumber
\end{equation}

and the observed duration of the $k^{\text{th}}$ process is $\delta_k = T - t_k$.  Because of the L\'{e}vy process assumption of Poisson arrivals, the expected commencement times are independent.  Then for the $k^{\text{th}}$ imagined process, the agent's estimate of the arrival rate is given by 

\[
\hat \lambda^{\mathcal{O}}_k = \frac{N_k}{\delta_k }
\]

and the estimator for the total arrival rate of the aggregated process can be written as 

\[
\hat \lambda^{\mathcal{O}}_{\Sigma} = \sum_{k=1}^K \frac{N_k}{\delta_k }
\]

Because commencement times (and thus observation periods) vary stochastically across processes $k$, the PKRE and non-speculative estimators for must be expressed as a summed series of terms. Given observation of all catastrophic events, the  expected losses due to catastrophic observed events are

\begin{equation}
\mathbf{J}^{\mathcal{O}}_{t} = \sum\limits_{k = 1}^{K^{\mathcal{O}}} {J^{\mathcal{O} }_{kt}} = \sum\limits_{k=1}^{K^{\mathcal{O}}} \sum\limits_{n=1}^{N_i(\delta) } Z_{kn,\tau} \implies     \mathcal{E}_{t}[\textbf{J}^{\mathcal{O}}] = \sum_{k=1}^{K^{\mathcal{O}}} \frac{\lambda^{\mathcal{O}}_k \xi^{\mathcal{O}}_k}{\mathcal{E}[\delta_k] }
    \nonumber
\end{equation}

and is seen by both speculative and non-speculative analysts.  

The speculative process for $\mathcal{E}_{\mathbf{R}}[\textbf{J}^{\mathcal{I}}]$ is staggered in a different way, by the unfolding of the dark speculation process itself.  The (intentionally) staggered sequence of speculation rounds allows us to reconceive of the ``error'' of the unspeculative analyst.  Until now, the $\pi_k$ have been treated as if drawn from a continuous, stationary distribution on $[0,1]$.  Let us now complicate the matter and focus upon $\pi_{rk}$ and allow the detection probabilities to vary within rounds. Suppose that $\pi_{rk}$ are increasing (or at least non-decreasing) in $r_k$, so that there is stochastic improvement in the non-speculative analyst's perception of possible catastrophic processes.  Let us posit an initial value of a series $\pi_1 : 0 < \pi_1 < \pi^{\max} < 1$ and then define $\pi^*_{rk} = \pi^{\max} - e^{-\psi (r_k-1)}\pi_1  $, where $\psi \in \Re^+$ is the coefficient of improvement.  Non-speculative perception of imagined catastrophes thus moves from an initial period to the non-speculative agent's maximum in a purely concave fashion.


Hence the non-speculative analyst observes 

\begin{equation}
    \mathcal{E}_{\text{nospec}}[\textbf{J}^{\mathcal{I}}] = \sum_{k=1}^{K^{\mathcal{I}}} \sum_{r_k=1}^{R_k} \frac{\pi^*_{rk} \hat\lambda_{rk} Z_{rk}}{\delta (r_k)}
    \nonumber
\end{equation}

and the following result is possible. \\

\begin{lemma}\label{lem:improvement} \textbf{Relative Benefits of Speculation Under Non-Speculative Improvement}: \textit{Given stochastic improvement in the non-speculative analyst as described by $\pi^*_k = \pi^{\max} - e^{-\psi (k-1)} \pi_1  $ $(\psi \in \Re^+)$, the benefits of speculation versus a non-speculative approach are greater early (for $r$ low), for lower initial state $\pi_1$, lower non-speculative ``peak'' $\pi^{\max}$ and lower coefficient of improvement $\psi$.}
\end{lemma}

Lemma \ref{lem:improvement} offers rather intuitive results, but establishes a useful baseline for counterfactuals.  If speculation is avoided, then the losses are greater unless the ``alternative'' policy has some degree of improvement, and even then, the benefits of speculation are likely greater in the near-term, implying possibly decreasing marginal returns to this institutionalized exercise.

Having elaborated a cursory mathematical expression of the framework, we now turn to more general considerations and extensions.

\section{Extensions: Hyperanxiety and Mitigation}

\subsection{Hyperanxiety for Red Line Events}

There has been considerable worry about the extreme downsides of frontier or generative AI, with one recent \textit{New York Times} article featuring a humanity-ending scenario induced by an AI prompt (\cite{witt2025}), analyzed by the AI pioneer Yoshua Bengio. More generally, \cite{acemoglu2024regulating} analyze a model in which ``adoption should be slower when ... damages from a disaster are large."  Since \cite{acemoglu2024regulating} examine the regulation of \textit{adoption}, however, some sense of the size of damages must be known or estimable ahead of time.  Estimation and learning about damages lies outside the framework of their model.  Yet the possibility that such a prospect would plausibly lead to \textit{ex ante} regulation delaying adoption, combined with the inevitable fact of uncertainty about estimates of the likelihood and severity of such events, raises the question: \textit{what if a social planner, regulator or insurer dramatically overestimates the probability or severity of a catastrophic event}?\footnote{We thank Guilherme Duarte for pushing us to analyze this question directly. What follows is a first cut.}  

Let us call a \textit{red line event} an event whose expected value would lead a regulator to prohibit or delay the adoption of AI.\footnote{Note that the net benefits of AI or of a given model are \textit{not} a subject of analysis in our model, though the estimates produced might be helpful in any such exercise. }   We are not any position to opine on whether Bengio or any other ``catastrophist'' is right or wrong.  The question becomes how such an event, and how inflated probability or severity estimates from such an event, would be handled in a dark speculation framework.

\subsubsection{Anxiety in Narrators and Underwriting Communities}

Let us postulate a threshold $\nu^* \in \Re^+$ for a red line event, satisfied when $\mathbf{X}^{\mathcal{Q}}_{t,R} > \nu^*$.  Let us imagine then that for the $k^{th}$ risk, the Agency produces an event in round $R$ \textit{and} that the Community produces hyperanxious probability and severity estimates for a maximal loss $\dot{Z}_R$, such that $\mathcal{E}_{R} [Z_k | \dot{Z}_{k,R}] = \dot{\lambda}^{\mathcal{I}} _k \cdot \dot{\xi}^{\mathcal{I}} _k$. The round's error is represented by $\beta_R$, which is the difference between the true probability and severity estimates and the erroneous ones ($\dot{\hat\lambda} \cdot \dot{\xi}^{\mathcal{I}} _k$), which are sufficiently divergent from the truth that they suffice to trigger a red line event: 

\[
\mathcal{E}_{R} [J^{\mathcal{I}}_k | \dot{Z}_{k,R}] = \lambda^{\mathcal{I}}_k \cdot \xi^{\mathcal{I}}_k + \beta_R 
\implies   \mathcal{E}_{R} \left( \mathbf{X}^{\mathcal{Q}}_{t,R}| \dot{Z}_{k,R} \right) > \nu^* \quad |\: \quad \{ \mathcal{E}_{R-1} \left( \mathbf{X}^{\mathcal{Q}}_{t,R-1} \right) \leq \nu^* \}
\]

If the non-speculative analyst is not so anxious, then with probability $1 - \pi_R$ she does not incorporate this wrongful estimate into her underwriting, hence the unspeculative analyst avoids over-anxious ascription of a red line event if

\[
1 - \pi >  \frac{ \dot{\hat\lambda}^{\mathcal{I}}_{k,R} \dot{Z}_{k,R} - \mathcal{E}_{R-1} \left( \mathbf{X}^{\mathcal{Q}}_{t,R-1} \right) }{R} - \nu^* 
\]

or if the probability of neglect is greater than the marginal contribution of the anxiously catastrophized event to triggering the red line threshold.  Such unwarranted red line event diagnoses are more likely earlier in the speculation process (lower $R$) as well as for more na\"{i}ve non-speculators (lower $\pi$).

Such is the ``pessimistic'' case.  But let us suppose that the Community is less anxious, or better at detecting the lower probability of an anxiously catastrophized event.  Put differently, let us suppose that the non-speculative agent -- or more properly, the agent who speculates but does not do so systematically and whose narratives are not met with skeptical probability and loss estimates from the Community -- is the na\"{i}ve one.  What can we say then about the potential benefits of dark speculation ?

\subsubsection{Anxious Narratives, Chill Underwriting Communities}

The TRIP Data Call scenarios generally involve cataclysmic events but they are localized and non-contagious.  In Section \ref{subsec:atlanta}, the car bomb hits Atlanta but the carbombers have not (in this scenario) developed and deployed other weapons.\footnote{See, however, \cite{TreasuryTRIP2024} for a case of large-scale radiation leakage from a terrorist attack on a commercial nuclear power reactor.}  Many scenarios of concern about AI involve large-scale, global threats.  There is good reason to take these threats seriously, but there is also the possibility that such possibilities might be inflated.  In what follows we \textit{presume} that the Agency produces a narrative that involves large-scale and massive human and non-human suffering, such as the bioweapon event in Section \ref{subsec:karnofsky} but at a much larger scale, or the detonation of one or more nuclear weapons.  We also presume that a na\"{i}ve observer considers this event non-trivially probable but that closer analysis would yield much less frequency, verging on zero.  How might the separation of informed underwriting from narrative generation help to reduce over-anxious catastrophizing?

Define by $\frac{\partial \hat\lambda^{\mathcal{I}}_{k,R}}{\partial \beta_R} < 0$ the relative precision of the underwriting Community.  The Community might erroneously estimate the arrival rate of the calamity, but the size of its error is decreasing in the calamity's magnitude.  The Community would ``begin'' with $\hat\lambda^{\mathcal{I}}_{k,R-1}$ but would adjust by virtue of analysis.\footnote{The adjustment process is exogenous to this analysis and would likely depend upon the experience, expertise and capacity of the Community.}  Then a simple condition for statistical improvement\footnote{By which we mean better estimates of parameters of interest (here $\lambda_k$).  Dark speculation might generate improvements to society's value via the revelation of mitigation strategies as well.} via dark speculation is

\begin{equation}
    \frac{\partial \hat\lambda^{\mathcal{I}}_{k,R}}{\partial \beta_R} > \frac{\partial}{\partial \dot{Z}_{k,R}} \left( \frac{ {\hat\lambda}^{\mathcal{I}}_{k,R-1} \dot{Z}_{k,R} - \mathcal{E}_{R-1} \left( \mathbf{X}^{\mathcal{Q}}_{t,R-1} \right) } { R} \right)= \frac{\hat\lambda^{\mathcal{I}}_{k,R-1}}{R}
\label{darkimprove}
\end{equation}

or if the Community's precision improves upon the marginal contribution of a calamity to the unadjusted posterior loss estimate.  This condition underscores a critical property of dark speculation as we have defined it:

\textit{Dark speculation separates the generation of catastrophic narratives from estimation of probabilities and loss severity, given a narrative.}  They are distinct processes (and plausibly independent) in the TRIP Data Call, and they are presumed independent in our analysis, though dependency structures might be modeled as extension to this framework.  Critically, this separation is what enables the Community of underwriters to approach a narrative skeptically and to assign it low frequency.

\subsection{Narratives with Mitigation}

So far the narratives produced assume some kind of ``success'' by malign actors, either in TRIP Data Call scenarios or in the bioweapon event considered in Section \ref{subsec:karnofsky}.  An important reason to engage in dark speculation, however, is that society will be better positioned to respond to malign intentions when equipped with a catastrophic scenario.  We term this process \textit{mitigation} and introduce a simplified definition of it here.\footnote{We thank Todd Lensman and Glen Weyl for suggesting game-theoretic representations of $\Omega^r$, such as in a game tree.  The particulars of this approach lie beyond the confines of this paper and represent a natural extension.}

\begin{definition}\label{def:mitigate}
\textbf{Mitigation}.  For any counterfactual LICAIN as defined in Definition~\ref{def:C-LICAIN}, a mitigation procedure has the following properties.
\begin{enumerate}
\item Given a pivotal happening $\breve\theta^{*}_{h,i} : \rho(\breve\theta^{*}_{h,i} = 1)$ such that $\exists \breve\chi_h (a^{\chi}_h): \rho(\breve\chi_h (a^{\chi}_h) = 1)$ or $ \exists \breve\zeta_h (a^{\zeta}_h): \rho(\breve\zeta_h (a^{\zeta}_h) = 1)$, there follows expected loss $ \mathcal{E}_r \left[ J^{\mathcal{I}}_{k,r} (\breve\theta^{*}_{h,i}) \right]$.
\item For an alternative happening $h^{\prime}$, such that $ \tilde\chi_{h^{\prime}} (\tilde{a}^{\chi}_{h^{\prime}}) : \rho(\tilde\chi_{h^{\prime}} (\tilde{a}^{\chi}_{h^{\prime}}) = 1) \implies \rho(\theta^*_{h,i} = 0)$ or $\exists \tilde\zeta_{h^{\prime}} (\tilde{a}^{\zeta}_{h^{\prime}}): \rho(\tilde\zeta_{h^{\prime}} (\tilde{a}^{\chi}_{h^{\prime}}) = 1) \implies \rho(\theta^*_{h,i} = 0)$, there follows expected loss $ \mathcal{E}_r \left[ J^{\mathcal{I} \prime}_{k,r} (\breve\theta^{*}_{h^{\prime},i}) \right] < \mathcal{E}_r \left[ J^{\mathcal{I}}_{k,r} (\breve\theta^{*}_{h,i}) \right]$

\item A \textit{Mitigator} is an agent or organization with round-specific aim 
\[
\underset{a_h \in \mathcal{A}} {\arg \min} \hspace{0.2cm}  \mathcal{E}_r \left[ J^{\mathcal{I}}_{k,r} \hspace{0.1cm} | \hspace{0.1cm}  \breve\theta^{*}_{h,i} \left( \breve\chi_h (a^{\chi}_h), \breve\zeta_h (a^{\zeta}_h) \right) \right]
\]

\item For at least one $\breve\theta^{*}_{h,i} : \rho(\breve\theta^{*}_{h,i} = 1)$, the Mitigator induces (by $\tilde{a}^{\chi}_{h^{\prime}}$, $\tilde{a}^{\zeta}_{h^{\prime}}$ or $\tilde{a}^{\chi \zeta}_{h^{\prime}}$ ) an alternative happening $h^{\prime}$ leading to  $ \mathcal{E}_r \left[ J^{\mathcal{I} \prime}_{k,r} (\breve\theta^{*}_{h^{\prime},i}) \right] $.

\end{enumerate}
\end{definition}

\subsubsection{Example: Mitigating a Bioweapon Disaster}

Applying Definition~\ref{def:mitigate} (Mitigation) to the bioweapon event in Section \ref{subsec:karnofsky}, let us suppose that a government or AI-firm establishes an if-then rule as follows.

\begin{itemize}
    \item a surveillance system for all $\gamma_2$ is established
    \item if any $\gamma_1$ queries a $\gamma_2$ according to happening $\theta_2$ in Section \ref{subsec:karnofsky}, then
    \item security agencies sweep known laboratories for supplies of any material used in $\theta_3$, and monitor attempts to use such laboratories or personnel with known access to such laboratories, and/or
    \item security agencies consider warnings to subway passengers or other public transit systems that would have been affected in $\theta_4$
\end{itemize} 

\subsubsection{Conditions and Benefits of Mitigation} 

Mitigation in the context of dark speculation depends upon at least two factors and perhaps more.

\begin{itemize}
    \item Mitigation presumes a framework in which some agent has imagined and narrated a catastrophe (the LICAIN as in Definition~\ref{def:LICAIN}).
    \item The catastrophic narrative had to admit of at least one counterfactual, one pivotal node,  upon which a Mitigator can act, according to Definition~\ref{def:C-LICAIN}.
    \item The process of underwriting may help identify how to allocate scarce law enforcement and security resources in a world of many possible attacks.
    \item Mitigated counterfactual loss narratives may present more realistic scenarios to the underwriting Community.
    
\end{itemize}

We believe that a compelling research agenda here is to deploy extensive form games and scenario analysis to examine mitigated catastrophic narratives.

\section{Optimal Policies for Pathological Speculation}

We now turn to the \textit{net} benefits of speculation under different scenarios.  We first state a general result and then turn to implementable rules based upon moments of the loss distribution being estimated.  \\

\subsection{The Statistical Value of Speculation: A Moment-Based Approach}

The benefits of dark speculation can be described in many ways, including the value of considering and developing mitigation strategies (which we examine below) and the value of learning.  We focus here on the statistical benefits, namely, the gains that come from better knowing and anticipating risk.  These depend upon assumed penalties for losses and how errors in estimation (for different moments of the loss distribution) map into values.  We let $m$ index moments, and write as $\Lambda_{m} (\mu_{m} ): \mu_{m} \mapsto \Re$ serve as a function mapping a moment $\mu_{m}$ to a decision-relevant variable.  (For instance, $\Lambda_{2} (\mu_{2} )$ maps the second moment of a distribution into its variance, and $\Lambda_{4} (\mu_{4} )$ converts the fourth moment into a leptokurtosis estimate. For any moment of $\mathbf{X}^{\mathcal{Q}}_{it}$, denote by $\Psi : \Delta \mathbf{X}^{\mathcal{Q}}_{it} \mapsto \Re^+$ a function (like a quadratic or absolute value) that maps differences into positive real values.  Further define by $F_M (\sum_k N , \sum_{k,n} Z) : \left\{\sum_k N , \sum_{k,n} Z \right\} \mapsto \Re $ a function that maps any moments of $N$ and $Z$ and any of their aggregates to a value dimension, and define by $\phi \in \Re$ a universal scalar multiplier.  We assume that $\Psi ( \cdot )$ is additively separable with respect to L\'{evy} components, such that $\Psi \left( \sum_{k=1}^{K^{\mathcal{I}}} \phi_k F_M \left( \lambda_k^{\mathcal{I}} , \xi_k^{\mathcal{I}} \right)  \right) = \Psi \left( \phi_1 F_M(\lambda^{\mathcal{I}}_1, \xi^{\mathcal{I}}_1) \right) + \Psi \left( \phi_2 F_M(\lambda^{\mathcal{I}}_2, \xi^{\mathcal{I}}_2) \right) + \dots \Psi \left( \phi_{K^{\mathcal{I}}} F_M(\lambda^{\mathcal{I}}_{K^{\mathcal{I}}}, \xi^{\mathcal{I}}_{K^{\mathcal{I}}} ) \right)  $. Further let $D_1 >0$ be the per-unit penalty from bias in estimating the rate of incurred losses and $D_2 >0$ be the per-unit loss associated with erroneously estimating the variance.\footnote{The function $\Psi ( \cdot )$ can be invariant across moments of $\mathbf{X}^{\mathcal{Q}}_{it}$ in our model because the series $D_1, D_2, ...$ varies across them.  It would be equivalent to set $D$ constant and allow $\Psi$ to vary by moments.}  Then the total statistical gains forfeited by not speculating would be

\begin{equation}
L_{k=K^{\mathcal{I}}} = \sum_{m} D_{m} \Psi \left( \Lambda_{m, \text{nospec}} \left( \sum_k N , \sum_{k,n} Z \right) - \Lambda_{m, r=K^{\mathcal{I}}} \left( \sum_k N , \sum_{k,n} Z \right) \right) 
\label{nospeclossgen}
\end{equation}

\subsubsection{Example: A Two-Moment Approach}

If we presume the first two moments are sufficient to describe the gains from speculation, then the total expected loss from not speculating is 

\begin{equation}
L^2_{k=K^{\mathcal{I}}} = D_1 \Psi \left( \sum_{k=1}^{K^{\mathcal{I}}} (\pi_k - 1) \hat\lambda_k^{\mathcal{I}} \cdot \hat\xi_k^{\mathcal{I}}  \right) + D_2 \Psi \left( \sum_{k=1}^{K^{\mathcal{I}}} (\pi_k - 1) \left( \hat\lambda_k^{\mathcal{I}} \sigma^2 + \hat\lambda^{\mathcal{I}} _k \hat\xi_k^{\mathcal{I}2}  \right) \right)
\label{nospecloss2m}
\end{equation}

\subsection{A General Policy}

For the general policy, let us define the benefits of \underline{m}itigation by $M_{k,r}: (0 < M_{k,r} < \infty) $, let $O_{k,r}: (0 < O_{k,r} < \infty) $ represent a generalized \underline{o}ption value, and let $\mathcal{F}_{k,r}$ represent the filtered probability space from dark speculation on risk $k$. Define the utility of speculation by $U ( \cdot): \{ L_k, M, O\} \mapsto \Re \hspace{0.2cm} s.t. -\infty < U < \infty$. Letting $s$ serve as a subindex for dark speculation rounds  $r$, the Government's dark speculation problem can be written as \\

\begin{equation}
V_s = \underset{r \geq s } {\sup} \hspace{0.2cm}  \mathcal{E}  \left[ U_s (\mathbf{J}_{k,s+1}^{\mathcal{I}} | \mathcal{F}^{\mathcal{I}}_{k,s})\right]    
\label{optimum}
\end{equation}

\textbf{Proposition 1: Optimal Speculation Policy for $k$th Risk}. \textit{For all rounds, there exists a unique solution to (\ref{optimum}) in which speculation is continued until an optimal stopping round $\tau^*$ and discontinued thereafter.  The optimal stopping round can be defined as} \\

\begin{equation}
\tau* =  {\inf} \hspace{0.2cm}  \{ \tau > 0: \left( V_s , U_s \right) \in \mathbb{Q}^*  \}    
\label{optstop}
\end{equation}

\textit{where $\mathbb{Q}^*$ is the stopping space and $\mathbb{C}^* = \Omega \hspace{0.1cm} \textbackslash \hspace{0.1cm} \mathbb{Q}^* $ is the continuation region.}

\subsection{Implementable Rules for the General Policy}

To produce implementable continuation rules, we need to specify the costs of the exercise and make further assumptions about the function $\Psi ( \cdot )$.  Given that rounds of speculation and happenings in the narrative are discrete variables, we proceed by first differences analysis.\footnote{We leave to future or other work the possibility of specifying the continuous ``effort'' of a pathological speculator.}  We examine two possibilities, first that speculation cost per round is constant and second that speculation cost varies by round with the size $H_r$ of the set of happenings $ \left\{ \theta \right\}$. In both cases we identify something of a continuation rule, namely that the policymaker should proceed with speculation given satisfaction of a given statistical condition, recognizing that ``stopping'' speculation is not irreversible. \\

\begin{corollary} \label{cor:optstationary}
\textbf{Optimal Speculation Policy Under Stationary Cost.}  \textit{Fix $c_{\textrm{spec}} = \overline{c}_{\textrm{spec}}$, and assume speculative measurement error as represented in \eqref{compoundvardiffnospec}.  Then given $K^{\mathcal{I}}$ components and $R_K$ periods of speculation previously for the $K^{th}$ component, and representing speculation benefits solely from the two-moment specification of statistical gains in  \eqref{nospecloss2m}, the policymaker should speculate iff }

\begin{multline}
c_{\textrm{write}} + \overline{c}_{\textrm{spec}} \leq  \Delta_{R,R-1} L^2_k + \Delta_{R,R-1} M + \Delta_{R,R-1} O \\
= \left( D_1 \Psi_{\mu_1} ( \mathbf{J}_{k,r}^{\mathcal{I}} | \mathcal{F}^{\mathcal{I}}_{k,r}) ) - D_1 \Psi_{\mu_1} ( \mathbf{J}_{k,r-1}^{\mathcal{I}} | \mathcal{F}^{\mathcal{I}}_{k,r-1})\right) \\
+ \left( D_2 \Psi_{\mu_2} ( \mathbf{J}_{k,r}^{\mathcal{I}} | \mathcal{F}^{\mathcal{I}}_{k,r}) ) - D_2 \Psi_{\mu_2} ( \mathbf{J}_{k,r-1}^{\mathcal{I}} | \mathcal{F}^{\mathcal{I}}_{k,r-1})\right) \\
+ ( M_R - M_{R-1} ) + ( O_R - O_{R-1} )
\label{specifconstant}
\end{multline}
\end{corollary}

Note a critical implication of Lemma 4 is that while \eqref{specifconstant} may counsel stopping the speculation process, as long as there are new imaginable processes that \textit{solus Deus scit}, optimal speculation and underwriting might occur infinitely.  If there is, instead, category learning where speculation about a particular carbombing or a particular bioweapon gives lessons about other catastrophic events (and thus stochastic processes) in the same class, then there is reason to think that there might, at some point, be decreasing marginal returns to the next series of speculation exercises.  Then again, since agentic AI and human creativity might, combined, concoct all sorts of new threats, perhaps the categorical learning will be limited, or new categories of risk (new $X^{\mathcal{SDS}}$ and hence new $X^{\mathcal{I}}$) will appear over time. 

Turning to variable speculation cost, we can imagine that the cost of the speculation exercise is a function of the size of the set of happenings $\left\{ \theta_r\right\}$ generated by the speculative exercise, which is $H_r$.  For each round of speculation, the analyst produces an $H_r$-tuple, and the round-specific cost is given by a concave function, for simplicity $c^r_s = \ln(1+H_r)$. \\

\begin{corollary} \label{cor:optvariable}
\textbf{Optimal Speculation Under Variable Speculation Cost, with Variable Speculation Benefit}.  \textit{Let us define the cost of speculation as $c^r_{\textrm{spec}} = \ln(1+H_r)$ and represent the benefit of more intensive narratives by $ \sigma^2_{\epsilon} (H) = \sigma^2_{\max} - e^{\eta (H_r-1)}  \sigma^2_{\min} $.\footnote{We use concave cost and improvement functions in this model, but it is easy to imagine convexity.  It might be that more happenings within each catastrophe complicate the narrative in highly non-linear ways, with interactions and dependencies among the different events, leading to greater and greater cost of narration (computation).}   Then given $R_k$ periods of speculation, speculation is optimal in the next $(R_K+1)^{\text{th}}$ period iff}

\begin{multline}
c_{\textrm{write}} + c^{R_K}_{\textrm{spec}} + \left( \ln(R_K+2) - \ln(R_K + 1) \right) \\
\leq  \Delta_{R,R-1} L^2_k + \Delta_{R,R-1} M + \Delta_{R,R-1} O \\
= \left( D_1 \Psi_{\mu_1} ( \mathbf{J}_{k,r}^{\mathcal{I}} | \mathcal{F}^{\mathcal{I}}_{k,r}) ) - D_1 \Psi_{\mu_1} ( \mathbf{J}_{k,r-1}^{\mathcal{I}} | \mathcal{F}^{\mathcal{I}}_{k,r-1})\right) \\
+ \left( D_2 \Psi_{\mu_2} ( \mathbf{J}_{k,r}^{\mathcal{I}} | \mathcal{F}^{\mathcal{I}}_{k,r} (H_r)) ) - D_2 \Psi_{\mu_2} ( \mathbf{J}_{k,r-1}^{\mathcal{I}} | \mathcal{F}^{\mathcal{I}}_{k,r-1} (H_{r-1}))\right) \\
+ ( M_R - M_{R-1} ) + ( O_R - O_{R-1} )
\label{specifvariable}
\end{multline}
\end{corollary}

Importantly, note that the benefits of more intensive speculation (driven by work reflected in $H$) show up only in the second-moment term, weighted by $D_2$.\\

\section{Conclusion: Benefits and Limitations of the Speculative Pathology Approach} 

\subsection{Regulation, Cost-Benefit Analysis, and the Limitations of AI-Risk Estimation Approaches}

The idea that certain kinds of systemic risk might be underestimated is a common worry in the realm of insurance for cyber threats (\cite{zeller2024accumulation}), terrorism (\cite{boardman2004known}) and climate catastrophes (\cite{charpentier2008insurability}).  Yet in cyberinsurance, which is one of the most rapidly-changing and innovative areas of risk analysis and clearly related to threats from AI, a recent review of methods (\cite{awiszus2023modeling}) reports two important limitations of the literature.  First, ``fewer papers have appeared that empirically analyze the respective severity distributions'' (9).  Second, even with more advanced actuarial methods that embed models of systemic risk using copula-based representations of stochastic processes, there is near-total reliance upon observational analysis, as modeling errors are ``mitigated by refining risk assessment procedures continuously based on expert input and evaluation of claims data'' (7).

Dark speculation may help.  It can be combined with a laissez-faire approach to regulating AI or with more stringent governance regimes.  It can assist insurance efforts in the frontier AI industry, but even without insurance regimes, dark speculation can also be useful to AI firms themselves as they consider the potential risks associated with their products.  As the epigraph suggests, moreover, a form of dark speculation is already occurring in the terrorism field.  

The \textit{combination} of narrative generation and underwriting exercises offers a clear and promising path forward.  Both would need to be institutionalized, and we think that investment in the speculation process likely lags well behind that of underwriting for AI, as there are already entrants into the AI insurance space (but see (\cite{schwarcz2025limits})).  

Moreover, dark speculation narratives may have public good properties.  An especially insightful, novel and realistic new narrative -- perhaps generated by a non-cooperative gaming scenario generation exercise -- introduces a sequence of events that had not been previously known ($ \left\{ SDS \right\}$) and has rendered it imaginable ($ \left\{ I \right\}$), perhaps for the first time.  Recent results from \cite{callander2024regulating} suggest that as entry into the AI field grows, a regulator may be less able to extract private information from firms, which would only worsen the undersupply of the kind of generalized risk information we discuss here.  Empirical research by \cite{strauss2025real} gives credence to this possibility, such that research on AI benefits (or profitability) is far outpacing (and at an ever higher rate) research on risks and costs. As underwriters wrestle with each catastrophic event narrative, and repeat the exercise, they might well learn how to better estimate (forecast) certain kinds of parameters.\footnote{Note that dynamic interactions among scenario-generators and underwriters are excluded from this model; there is no ``learning-by-doing,'' just statistical inference on frequency and severity parameters, conditional on a generated narrative.}

Finally, in the event that some form of regulation develops for AI, both cost-benefit analysis of AI deployment and cost-benefit analysis of the regulation itself will be critical (\cite{acemoglu2024regulating}, \cite{callander2024regulating}).  Even if only adverse events were observed and not imagined, the underwriting exercise discussed here would be useful.  And we believe that the best way to falsify the hypothesis of real value in dark speculation is to engage in the process and learn if in fact it conveys little utility for management or policymaking.

\subsection{Results and Limitations of the Stochastic Process Model}

\subsubsection{Lessons from the Model }

Corollaries \ref{cor:optstationary} and \ref{cor:optvariable} convey general results in terms of a range of variables.  They also point to some important principles for optimization.  

\begin{itemize}
\item The rounds of speculation are presumed independent.  Repeated speculation on the same category of processes might introduce efficiencies (Corollary \ref{cor:optvariable}), but might also deprive speculators of the benefits of heuristic diversity (\cite{hong2001problem}), or the benefits that come from different ``takes'' on a problem that are irreducible to those used previously.
\item Important costs in our model -- the institutional cost of the speculation enterprise and that of the underwriting call -- are, in a sense, ``endogenous'' to policymaking.  A public authority, a non-profit organization or an investment firm might establish a scenario-generating institute -- like MITRE, RAND, or the Georgetown Wargaming Society -- and one might expect costs to decline based upon learning-by-doing, scale economies and operational efficiencies (Corollary \ref{cor:optvariable}).
\item Once dark speculation has commenced, the value of speculation depends heavily upon expectations of the \textit{next} stochastic catastrophic loss process that has not yet been imagined (Corollary \ref{cor:optstationary}).  We presume stable expectations over these variables, yet this assumption may be violated easily, thereby complicating policymaking.
\item If firms, policymakers or insurers were to consider institutionalized speculation, they might ask how well a counterfactual, non-speculating status quo considers the character, frequency and losses associated with catastrophic events (Lemma \ref{lem:improvement}).  We model this dynamic probabilistically but, outside of settings where dark speculation is institutionalized, assigning even an appreciable probability to detection of imaginable risks may be too optimistic. 
\end{itemize} 

\subsubsection{Limitations of the Model and Idea, Including Questions for Further Research}

Importantly, the framework here ignores, by definition, factors that should be considered in further efforts and in building any kind of dark speculation institution.

\begin{itemize}
\item We have not considered one of the most important benefits of speculation, namely that engaging in speculation might provide policymakers and firms with mitigation ``tools'' to combat the catastrophic risk.  Of course, malicious actors might also learn from such a process or institutionalize it themselves.
\item While we model catastrophes arriving as memoryless Poisson jumps, real-world events are more likely to be self-exciting——a series of attacks may happen consecutively after a new exploit is discovered or capability threshold is reached. Even across different risk families, risks can be correlated and cascading \citep{kulveit_gradual_2025}——damage to power grids and transportation, for example, would have significant downstream consequences for financial markets. However, our core insight that risk evaluation improves with speculation remains valid with correlated risks, and modifications to our model can account for these correlations. Furthermore, the iterative, speculative process we propose is appropriate for managing cross-sector contagion, both via scenario generation and pricing.
\item The arrival of risks may also be strategic——misaligned AI systems may alter their behavior based on what human overseers have speculated and are prepared to detect if attempted. Misaligned AI systems may, for example, be more likely to take harmful actions that humans have not yet imagined. This possibility introduces a new cost to speculation \citep{stenseke_counter-productivity_2025}. From an institutional design perspective, it suggests that some speculation should remain private and unknown to misaligned AI systems. This, however, may create frictions that limit information sharing and aggregation about scenarios.
\item Another consideration is that scenario generation may be biased by availability (e.g., similarity to already imagined or observed scenarios), towards overdramatic or underdramatic events, or towards risks better understood by the speculating community (e.g., cultural or domain-specific biases). Even biased speculation outperforms no speculation on risk evaluation if it is positively generating new scenarios. However, blindspots in scenario generation may prevent risk estimation from approaching the true risk. Addressing this problem is, in part, a matter of institutional design, including employing diverse speculation teams and scrutinizing or auditing scenario generation strategies. However, some scenarios may remain outside a diversified scenario market, and we treat this residual $\mathcal{SDS}$ mass as exogenous.
\item We have not considered the possibility that if an institutional ``machine'' of dark speculation (like an institute) were up and running, stopping it might result in irreversible losses due to forfeited expertise and the diversion of human, physical and financial capital to other ends.
\item We have not considered, relatedly, the risk appetite of the policymaker (the framework here has been constructed under risk neutrality).
\item Any regulatory regime that deployed speculative pathology would have to avoid implementing the most na\"{i}ve decision rules.  Just because a war-gaming or speculative pathology exercise can produce a horrific imagined result -- the end of the world -- should not imply that the most restrictive regulatory response should be adopted.\footnote{One can imagine formalizing such a na\"{i}ve aproach in assigning infinite or massive loss values to a mere singleton in the imaginable adverse event set $\left\{ \mathcal{I} \right\}$, at which point the risks of foundation models might greatly outweigh any benefit from their existence.}  Any speculative exercise that included the worst possible scenario would also need to consider humanity's likely best response in addition to regulatory options.
\end{itemize}

Dark speculation already exists at a small scale in several domains and we believe that researchers and policymakers could institutionalize the process more generally.  Dedicated efforts in both scenario generation (or wargaming) and high-complexity insurance modeling can and should inform one another, but we think this becomes likely only when the public good of new, rich and realistic narratives are produced at scale.

\newpage

\section{Appendix 1: Proofs}

\textbf{Proof of Lemma 1}. The multi-component jump processes can be described by moment-generating functions, which for the imagined set $\left\{ \mathcal{I} \right\}$ is given, for the \textit{k}th component, by

\begin{equation}
M^{\mathcal{I}}_{J_{k,t}} (s) = \text{exp} \left( {\hat\lambda_k^{\mathcal{I}}} t \left[ M^{\mathcal{I}}_{Z_{k,t}} (s) - 1\right] \right) 
\label{mgfcomponent}
\end{equation}

and for the sum $\mathbf{J}^{\mathcal{I}}_{t}$,

\begin{multline} M^{ \mathcal{I}}_{\Sigma} (s) = \prod_{k=1}^{K^{\mathcal{I}}} \text{exp}\left( \hat\lambda^{\mathcal{I}}_k t \left( M^{\mathcal{I}}_{Z_{k,t}} (s) - 1 \right) \right)  \\
\implies M^{\prime \mathcal{I}}_{\Sigma} (s)  =  \sum_{k=1}^{K^{\mathcal{I}}} \hat\lambda^{\mathcal{I}}_k t M^{\prime \mathcal{I}}_{Z_{k,t}} (s) \left(\text{exp}\left( \hat\lambda^{\mathcal{I}}_k t \left( M^{\mathcal{I}}_{Z_{ik,t}} (s) - 1 \right) \right) \right)
\label{mgfsum}
\end{multline}

Taking the derivative of each $M^{\mathcal{I}}_{J_{k,t}} (s)$ with respect to $s$ and setting the derivative equal to zero, we get, for each component $X^{\mathcal{Q}}_{kt}$, $M^{\prime \mathcal{I}}_{J_{k,t}} (0) = t \lambda^{\mathcal{I}}_k \xi^{\mathcal{I}}_k $ and $M^{\prime \mathcal{I}}_{\Sigma} (0) = \sum_{k=1}^{K^{\mathcal{I}}} t \hat\lambda^{\mathcal{I}}_k \hat\xi^{\mathcal{I}}_k $ . \\

The non-speculating agent observes only a fraction of each process $\pi_k$, hence for a number of rounds equal to $r = K^{\mathcal{I}}$, the MGF can be written as

\begin{equation}
M^{\mathcal{I}}_{\Sigma, \text{nospec}} = \text{exp} \sum_{k=1}^{K^{\mathcal{I}}} \left( \pi_k \hat\lambda^{\mathcal{I}}_k \left( M^{\mathcal{I}}_{Z_{k}} (s) - 1 \right) \right) 
\label{MGFnospec}
\end{equation}

The relation \eqref{MGFnospec} shows that the ``failure'' to speculate enters linearly into each moment of the estimated loss distribution, affecting not only means but estimates of tail properties, assuming these can be expressed as moments of a stable distribution.  \\

Similarly, for each component $X^{\mathcal{Q}}_{kt}$, $M^{\prime \mathcal{I}}_{\text{nospec}} (0) = t \pi_k \hat\lambda^{\mathcal{I}}_k \hat\xi^{\mathcal{I}}_k $ and $M^{\prime \mathcal{I}}_{\Sigma, \text{nospec}} (0) = \sum_{k=1}^{K^{\mathcal{I}}} t \pi_k \hat\lambda^{\mathcal{I}}_k \hat\xi^{\mathcal{I}}_k $ . From here the total bias of the non-speculative agent as compared to the PKRE can be written as the difference between these two conditional means net of time observed, or

\begin{equation}
\beta_{\Sigma, \text{nospec}} (\mathcal{E}_{\mathbf{R}} [\overline{\mathbf{X}}^{\mathcal{Q}}_{it}]) = \sum_{k=1}^{K^{\mathcal{I}}} (\pi_k - 1) \hat\lambda_k^{\mathcal{I}} \cdot \hat\xi_k^{\mathcal{I}}  
\label{compoundbiasnospec}
\end{equation}

$\square$ \\

\textbf{Proof of Lemma 2}. Proceeding again from the MGFs \eqref{mgfcomponent} and \eqref{mgfsum}, we calculate the second moment $M^{\prime \prime \mathcal{I}}_{\Sigma} (0)$ and subtract $(M^{\prime \mathcal{I}}_{\Sigma} (0))^2$, to get

\begin{equation}
\text{Var} (\mathcal{E}_{\mathbf{R}} [\overline{\mathbf{X}}^{\mathcal{Q}}_{t}]) = \sum_{k=1}^{K^{\mathcal{I}}}  \hat\lambda_k^{\mathcal{I}} \left( \sigma_k^2 (Z^{\mathcal{I}}_k) + \hat\xi_k^{\mathcal{I}2} \right)
\label{varcompound}
\end{equation}

An identical calculation for $\text{Var} (\mathcal{E}_{\text{nospec}} [\overline{\mathbf{X}}^{\mathcal{Q}}_{t}]) $ gives the summed variances in \eqref{varcompound}, each weighted by the probability of process ignorance $(1 - \pi_k)$. $\square$ \\

\textbf{Proof of Lemma 3}.  We begin with the $k^{\text{th}}$ component.  Because $r_k$ is an integer variable, we use the method of first differences.  We can express the PKRE at the $R_k^{\text{th}}$ period by writing out its penultimate two-round movement, as follows

\begin{equation}
 \mathcal{E}_{\textbf{R}}[\textbf{J}^{\mathcal{I}}] = \sum_{r_k=1}^{R_k -2} \frac{\hat\lambda_r Z_r}{\delta_r} + \frac{\hat\lambda_{R_k -1} Z_{R_k-1}}{\delta_{R_k - 1}} + \frac{\hat\lambda_{R_k} Z_{R_k}}{\delta_{R_k}}
 \nonumber
\end{equation}

For the non-speculative estimate, the same exercise gives

\begin{equation}
 \mathcal{E}_{\text{nospec, improve}}[\textbf{J}^{\mathcal{I}}] = \sum_{r_k=1}^{R_k -2} \frac{\pi^*_r \hat\lambda_r Z_r}{\delta_r} + \frac{ (\pi^{\max} - e^{-(R-1) \psi}\pi_1) \hat\lambda_{R_k -1} Z_{R_k-1}}{\delta_{R_k -1 }} + \frac{ (\pi^{\max} - e^{-R \psi}\pi_1)  \hat\lambda_{R_k } Z_{R_k}}{\delta_{R_k}}
 \nonumber
\end{equation}

Taking expectations, the movement of $\mathcal{E}_{\text{nospec}}[\textbf{J}^{\mathcal{I}}] $ between the $(R-1)^{\text{th}}$ and $R^{\text{th}}$ periods is $(\pi^{\max} - \pi_1) (e^{-k \psi} - e^{-(k-1) \psi})  \xi_K (\mathcal{E} [\tau_K])^{-1}
$ and the difference in one-period movement between the PKRE and the improving non-speculative agent is 

\begin{equation}
\Delta_{R_k, R_k-1} \left( \mathcal{E}_{\text{nospec}}[\textbf{J}^{\mathcal{I}}], \mathcal{E}_{R_k}[\textbf{J}^{\mathcal{I}}] \right) = \frac{ \left( 1 - (\pi^{\max} - e^{-(R_k-1) \psi}\pi_1 )    \right)  \hat\lambda_{R_k } \hat\xi_{R_k}} {\mathcal{E} [\delta_{R_k}]}
\label{DeltaDiff}
\end{equation}

Upon inspection $\Delta_{R_k, R_k-1} \left( \mathcal{E}_{\text{nospec}}[\textbf{J}^{\mathcal{I}}], \mathcal{E}_{R_k}[\textbf{J}^{\mathcal{I}}] \right)$ is decreasing in $R_k$ and $\psi$ .  For $\pi^{\max}$, simple substitute $\pi^*$ for $\pi$ in \eqref{compoundbiasnospec}.  For $\pi_1$, note that, fixing $\pi^{\max}$, $\Delta_{R_k, R_k-1} \left( \mathcal{E}_{\text{nospec}}[\textbf{J}^{\mathcal{I}}], \mathcal{E}_{R_k}[\textbf{J}^{\mathcal{I}}] \right)$ is decreasing in $\pi_1$.  $\square$ \\

\textbf{Proof of Proposition 1}.  We first demonstrate bounded stochastic value.  By the definition of $Z_k$, the following constraint holds.

\begin{equation}
\mathcal{E} \left( \underset{s \leq r \leq S } {\sup} \hspace{0.2cm}   \left[ \mathbf{J}_{k,r+1}^{\mathcal{I}} | \mathcal{F}^{\mathcal{I}}_{k,r}\right] \right) < \infty
\nonumber    
\end{equation}

hence \\

\begin{equation}
\mathcal{E} \left( \underset{r \geq s } {\sup} \hspace{0.1cm} \left[ U_s (\mathbf{J}_{k,s+1}^{\mathcal{I}} | \mathcal{F}^{\mathcal{I}}_{k,s})\right] \right) < \infty
\nonumber    
\end{equation}

Any solution to (\ref{optimum}) must satisfy a dynamic programming relation (\cite{bertsekas1996stochastic}: Corollary 9.12.1), which for Borel-measurable spaces and an infinite horizon can be written as 

\begin{equation}
\underset{r \geq s } {\sup} \hspace{0.1cm} \left[ u_s (\mathbf{J}_{k,s}^{\mathcal{I}} | \mathcal{F}^{\mathcal{I}}_{k,s-1}) + \rho \int U^*_{s+1} (\mathbf{J}_{k,s+1}^{\mathcal{I}} | \mathcal{F}^{\mathcal{I}}_{k,s}) \hspace{0.1cm} d\textbf{J}_k^{\mathcal{I}} \right]  < \infty
\label{bellman}    
\end{equation}

where $\rho: (0 < \rho < 1)$ is a constant discount factor, $u_s$ is the per-period ``flow'' value and $U_s^*(\cdot)$ presumes the optimality of actions taken in the next period (and by recursion, all periods thereafter).

Then by \cite{peskir2006optimal}, Theorem 1.4, an optimal stopping solution exists and the stopping time is either finite and unique or infinite. Because stochastic integrals are Borel-measurable and by the monotone convergence theorem (\cite{bertsekas1996stochastic}: Lemma 7.11), since the relationship in (\ref{bellman}) holds for $s = r$, it holds $\forall s$. $\square$ \\

\textbf{Proof of Corollary 1.1}.  Because the additive separability of $\Psi(\cdot)$ applies across rounds $r_k$ as well as across components $k$, the moment-based loss function $L_{\mathbf{R}}$ can, upon the $R_k{\text{-th}}$ round of speculation, be written as a sum of the $K^{\mathcal{I}}-1$ other components and, for the $K$th component, the sum of the $R_K - 1$ previous rounds and the last $R_K$th round.  

\begin{multline}
L_{\mathbf{R}} = D_1 \Psi \left( \sum_{k=1}^{K - 1} (\pi_k - 1) \hat\lambda_k^{\mathcal{I}} \cdot \hat\xi_k^{\mathcal{I}}  \right) + D_2 \Psi \left( \sum_{k=1}^{K - 1} (\pi_k - 1) \left( \hat\lambda_k^{\mathcal{I}} \sigma^2 + \hat\lambda^{\mathcal{I}} _k \hat\xi_k^{\mathcal{I}2}  \right) \right) + \\
D_1 \Psi \left( \frac{\sum_{r_K=1}^{R_K- 1} (\pi_r - 1) \hat\lambda_{rK}^{\mathcal{I}} \cdot \hat\xi_{rK}^{\mathcal{I}}}{R_K - 1}  \right) + D_2 \Psi \left( \frac{\sum_{r_k=1}^{R_K - 1} (\pi_r - 1) \left( \hat\lambda_{rK}^{\mathcal{I}} \sigma^2 + \hat\lambda^{\mathcal{I}} _{rK} \hat\xi_{rK}^{\mathcal{I}2}  \right)}{R_K-1} \right) \\
+ D_1 \Psi \left( (\pi_{R_K} - 1) \hat\lambda_{R,K}^{\mathcal{I}} \cdot \hat\xi_{R,K}^{\mathcal{I}}  \right) 
+ D_2 \Psi \left( (\pi_{R_K} - 1) \left( \hat\lambda^{\mathcal{I}}_K  \sigma^2 + \hat\lambda^{\mathcal{I}} _{RK} \hat\xi_{RK}^{\mathcal{I}2} - \hat\lambda^{\mathcal{I}} _{RK} \sigma^{\mathcal{I} 2}_{\epsilon, K} \right) \right)
\nonumber
\end{multline}

Then $\forall k$ and taking period-expectations over the intensity and jump-size distribution, and with ignorance probabilities $\pi_k$ known and deterministic, the next round's incremental addition to the two-moment-based statistical loss function is

\begin{multline}
D_1 \Psi \left( (\pi_{K,R} - 1) \mathcal{E}_{\mathbf{R}}[\lambda_{K,R}^{\mathcal{I}}] \cdot \mathcal{E}_{\mathbf{R}}[\xi_{K,R}^{\mathcal{I}}] - (\pi_{K,R-1} - 1) \mathcal{E}_{\mathbf{R}}[\lambda_{K,R-1}^{\mathcal{I}}] \cdot \mathcal{E}_{\mathbf{R}}[\xi_{K,R-1}^{\mathcal{I}}]  \right) \\
+ D_2 \Psi \left( (\pi_{k^{\prime} + 1} - 1) \left( \mathcal{E}_{\mathbf{R}}[\lambda^{\mathcal{I}}_{K,R+1}]  \sigma^2 + \mathcal{E}_{\mathbf{R}}[\lambda^{\mathcal{I}} _{K,R}] \mathcal{E}_{\mathbf{R}}[\xi_{K,R}^{\mathcal{I}2}] - \mathcal{E}_{\mathbf{R}}[\lambda^{\mathcal{I}} _{K,R}] \sigma^{\mathcal{I} 2}_{\epsilon, {K,R}} \right) \right)- \\
D_2 \Psi \left( (\pi_{k^{\prime} + 1} - 1) \left( \mathcal{E}_{\mathbf{R}}[\lambda^{\mathcal{I}}_{K,R-1}]  \sigma^2 + \mathcal{E}_{\mathbf{R}}[\lambda^{\mathcal{I}} _{K,R-1}] \mathcal{E}_{\mathbf{R}}[\xi_{K,R-1}^{\mathcal{I}2}] - \mathcal{E}_{\mathbf{R}}[\lambda^{\mathcal{I}} _{K,R-1}] \sigma^{\mathcal{I} 2}_{\epsilon, {K,R-1}} \right)\right) \\
= \left( D_1 \Psi_{\mu_1} ( \mathbf{J}_{k,r}^{\mathcal{I}} | \mathcal{F}^{\mathcal{I}}_{k,r}) ) - D_1 \Psi_{\mu_1} ( \mathbf{J}_{k,r-1}^{\mathcal{I}} | \mathcal{F}^{\mathcal{I}}_{k,r-1})\right) \\
+ \left( D_2 \Psi_{\mu_2} ( \mathbf{J}_{k,r}^{\mathcal{I}} | \mathcal{F}^{\mathcal{I}}_{k,r}) ) - D_2 \Psi_{\mu_2} ( \mathbf{J}_{k,r-1}^{\mathcal{I}} | \mathcal{F}^{\mathcal{I}}_{k,r-1})\right)\nonumber
\end{multline}

Because $M$ and $O$ are separable from $L^{\mu}_{k,R}$, these terms enter linearly into the continuation rule and the inequality follows. $\square$ \\

\bibliography{references.bib}

\end{document}